\documentclass[a4paper,fleqn]{sty/cas-dc}

\usepackage[numbers]{natbib}
\usepackage{epsfig}
\usepackage[inline]{enumitem}
\usepackage{caption}
\usepackage{stackengine}
\usepackage{subcaption}
\usepackage{balance}
\usepackage{xspace}

\setlist[itemize]{noitemsep, topsep=0pt}
\setlist[enumerate]{noitemsep, topsep=0pt}

\newcommand{\ie}{\textit{i.e.}, }
\newcommand{\eg}{\textit{e.g.}, }
\newcommand{\etc}{\emph{etc.\xspace}}
\newcommand{\etal}{\emph{et al.\xspace}}

\newcommand{\HEVC}{\emph{High Efficiency Video Coding }}
\newcommand{\AVC}{\emph{Advanced Video Coding }}
\newcommand{\VVC}{\emph{Versatile Video Coding }}
\newcommand{\AVONE}{\emph{AOMedia Video 1 }}
\newcommand{\ekrem}[1]{\textcolor{black}{#1}}
\newcommand{\ct}[1]{\textcolor{black}{#1}}
\newcommand{\hadi}[1]{\textcolor{black}{#1}}
\newcommand{\revision}[1]{\textcolor{black}{#1}}

\graphicspath{{figs/}}
\newcolumntype{?}{!{\vrule width 1pt}}
\def\tsc#1{\csdef{#1}{\textsc{\lowercase{#1}}\xspace}}
\tsc{WGM}
\tsc{QE}
\tsc{EP}
\tsc{PMS}
\tsc{BEC}
\tsc{DE}

\begin{document}
\let\WriteBookmarks\relax
\def\floatpagepagefraction{1}
\def\textpagefraction{.001}
\shorttitle{CTU Depth Decision Algorithms for HEVC: A Survey}
\shortauthors{Ekrem Çetinkaya, Hadi Amirpour et~al.}

\title [mode = title]{CTU Depth Decision Algorithms for HEVC:\\ A Survey}                      

\author[1,3]{Ekrem Çetinkaya}[type=editor,
                        auid=,bioid=,
                        prefix=,
                        role=,
                        orcid=0000-0002-6084-6249]
\author[1,3]{Hadi Amirpour}[orcid=0000-0001-9853-1720]
\author[1,2]{Mohammad Ghanbari}[orcid=0000-0002-5482-8378]
\author[1]{Christian Timmerer}[orcid=0000-0002-0031-5243]

\address[1]{Christian Doppler Laboratory ATHENA, Alpen-Adria-Universität, Klagenfurt, Austria}
\address[2]{School of Computer Science and Electronic Engineering, University of Essex, Colchester, UK}
\address[3]{These authors contributed equally to this work.}


\begin{abstract}
High Efficiency Video Coding (HEVC) surpasses its predecessors in encoding efficiency by introducing new coding tools at the cost of an increased encoding time-complexity. The Coding Tree Unit (CTU) is the main building block used in HEVC. In the HEVC standard, frames are divided into CTUs with the predetermined size of up to $64\times64$ pixels. Each CTU is then divided recursively into a number of equally sized square areas, known as Coding Units (CUs). Although this diversity of frame partitioning increases encoding efficiency, it also causes an increase in the time complexity due to the increased number of ways to find the optimal partitioning. To address this complexity, numerous algorithms have been proposed to eliminate unnecessary searches during partitioning CTUs by exploiting the correlation in the video. In this paper, existing CTU depth decision algorithms for HEVC are surveyed. These algorithms are categorized into two groups, namely statistics and machine learning approaches. Statistics approaches are further subdivided into neighboring and inherent approaches. Neighboring approaches exploit the similarity between adjacent CTUs to limit the depth range of the current CTU, while inherent approaches use only the available information within the current CTU. Machine learning approaches try to extract and exploit similarities implicitly. Traditional methods like support vector machines or random forests use manually selected features, while recently proposed deep learning methods extract features during training. Finally, this paper discusses extending these methods to more recent video coding formats such as \VVC (VVC) and \AVONE (AV1).

\end{abstract}

\maketitle

\section{Introduction}

Video streaming has become an essential part of today's Internet traffic. The majority of the global Internet traffic consists of video (75\% in 2017), and its share is expected to grow in the future (82\% by 2022)~\cite{CiscoForecast}. This steep increase in video traffic has created challenges in several blocks of the video streaming solutions that need to be addressed. The building blocks of the video streaming can be expressed as content provisioning, content delivery, and content consumption. In this survey, we will focus on the content provisioning part (\ie video coding) of the video streaming scheme, while for content delivery and consumption parts, we refer to Bentaleb~\etal~\cite{ABRSurvey}.

A video is \hadi{a} sequence of images with redundant information. Spatial and temporal redundancy in videos can be exploited to reduce their size. \ekrem{This process requires efficient coding tools, since finding similarities is the key for exploiting redundancy, and \hadi{the} vast amount of information in video\hadi{s} makes this process difficult. 
Also, as the framerate, resolution\hadi{,} and bit depth of video content increases, the amount of information available in the video increases significantly, making the \ct{reduction} in the amount of redundancy \ct{more and more} important.} This led to the need to develop more advanced video encoders beyond the existing \AVC \hadi{(AVC)} standard~\cite{AVC}. \HEVC (HEVC)~\cite{HEVC} is the successor of AVC that has been developed by the Joint Collaborative Team on Video Coding (JCT-VC) which has improved the existing encoding tools and introduced new ones to increase encoding efficiency. Compared to \ct{AVC}, HEVC can reduce the bitrate of video by 50\%~\cite{HEVC}. 

Modern video encoders adopt a block-based structure for motion compensation to improve encoding efficiency. In \hadi{the} block-based structure, frames are divided into several smaller blocks that vary in size depending on the complexity of the content. These blocks are later used in the motion compensation part of the encoder\hadi{,} in which the encoder tries to predict the block using the best-matched block in the current frame (Intra) or in the previously encoded frames (Inter) based on the motion information. After a block is predicted, the residual error information is transformed into transform blocks, and they are entropy encoded. 

Each video codec uses (slightly) different block structures. Older video coding standards like MPEG-2~\cite{MPEG2} use a fixed block size of $16\times16$ \hadi{pixels}, while their transform blocks have a size of $8\times8$ samples.
The more recent AVC standard introduced a more flexible block structure.

In AVC, frames are divided into macroblocks of varying sizes up to $16\times16$ pixels\hadi{,} and each macroblock can \hadi{be} further partitioned into variable sizes. For intra prediction, \hadi{sub-macroblock sizes of} $16\times16$, $8\times8$, and $4\times4$ \hadi{pixels} are allowed. For inter prediction, \hadi{the sub-macroblock sizes of}  $16\times16$, $16\times8$, $8\times16$, and $8\times8$\hadi {pixels} are searched and a motion vector is assigned to each sub-macroblock. 
When $8\times8$ is an optimal sub-macroblock size candidate, $8\times4$, $4\times8$, and $4\times4$ sub-macroblock sizes are also checked for each $8\times8$ sub-macroblock. Depending on the size selected, one of two $8\times8$ or $4\times4$ transform blocks will eventually be selected~\cite{HEVCBlockPartitioning}. 

Using larger blocks to exploit \hadi{more} spatial redundancy can increase \hadi{the} efficiency and flexibility of the encoder\hadi{,} especially for higher resolution videos~\cite{HEVCBlockPartitioning, Ma2007HighdefinitionVC}. HEVC introduces a new block partitioning structure called Coding Tree Unit (CTU), which can vary in size from $8\times8$ to $64\times64$ pixels~\cite{Wien:2014:HEV:2685406}. Despite the increased efficiency and flexibility of the HEVC, using the CTU as a building block also leads to significant complexity in encoding time, making the use of HEVC a challenging task for several applications, \eg live streaming. To cope with this increased complexity, many algorithms have been proposed to reduce the process of rate-distortion optimization (RDO) by eliminating unnecessary searches for optimal CTU partitioning using several available sources of information. 

\begin{figure*}[pos=t]
\captionsetup{justification=centering,
              singlelinecheck=false,
              font=sf, labelfont=bf} 
\centering
\includegraphics[width=0.9\textwidth]{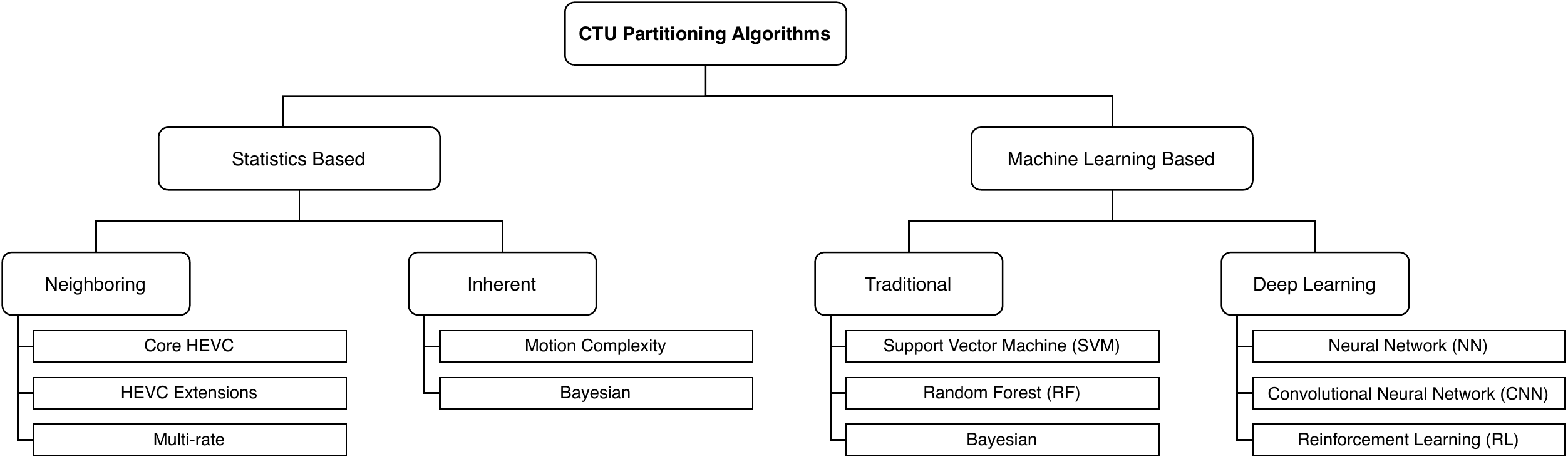}
\caption{Classification of CTU partitioning algortihms.}
\label{fig:cuapproaches}
\end{figure*}

This paper provides a comprehensive study of CTU partitioning algorithms. We give detailed information about \hadi{the} CTU structure of HEVC in Section~\ref{sec:ctu}. \ekrem{We then classify the existing methods and mainly focus on the CU depth decision algorithms for HEVC. We categorize existing approaches into statistics based and machine learning (ML) based methods. The former benefits from \hadi{the} statistical correlation in the video, while the latter uses machine learning methods to extract correlation. We present these approaches in Section~\ref{sec:statistic} and Section~\ref{sec:Machine}\hadi{,} respectively. Fig.~\ref{fig:cuapproaches} shows the broad categorization for CU depth decision algorithms used in this paper. In Section~\ref{sec:discussion}, we discuss the overall findings and possible future directions \ct{including emerging coding formats such as \VVC (VVC) and \AVONE (AV1)}\hadi{. Finally, we}  conclude the paper in Section~\ref{sec:conclusion}}.


\section{Overview of HEVC CTU partitioning}
\label{sec:ctu}

In HEVC, frames are divided into tiles or slices\hadi{,} which are further divided into non-overlapped CTUs. Each CTU can then be split into several square regions of equal sizes, known as coding units (CUs), using a quad-tree structure. 

\subsection{Coding Tree Unit (CTU)}


\begin{figure*}[pos=t]
\captionsetup{justification=centering,
              singlelinecheck=false,
              font=sf, labelfont=bf} 
\centering
\includegraphics[width=0.6\textwidth]{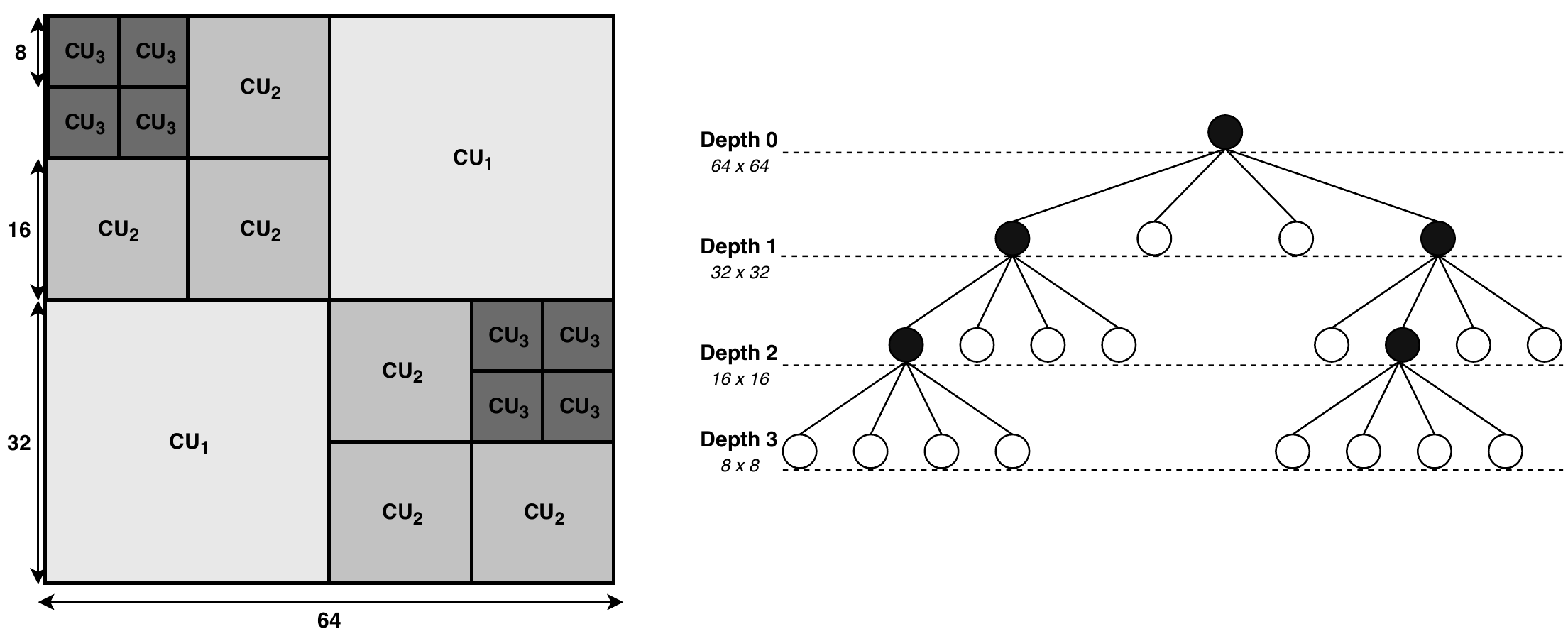}
\caption{CTU partitioning of HEVC.}
\label{fig:ctu}
\end{figure*}
\begin{figure*}[pos=t]
\captionsetup{justification=raggedright,
              singlelinecheck=false,
              font=sf, labelfont=bf} 
\centering
\includegraphics[width=0.9\textwidth]{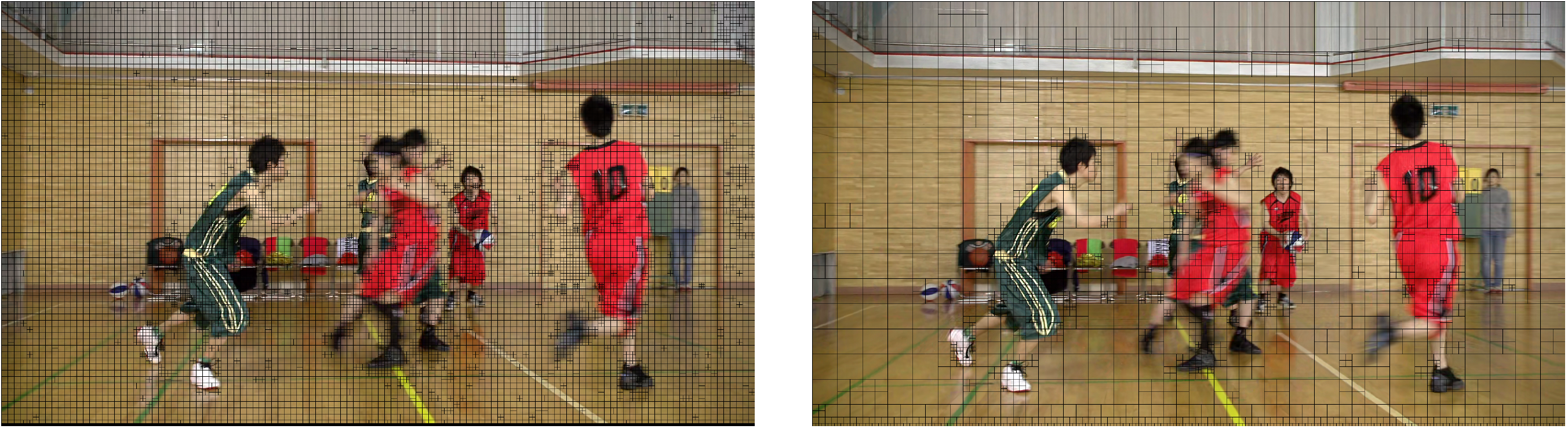}
\caption{Comparison of block partitioning between AVC and HEVC for the 100th frame of the BasketballDrive sequence. Both frames have been encoded at the same bitrate.}
\label{fig:avchevc}
\end{figure*}

\ekrem{HEVC uses \hadi{a} \textit{quad-tree} structure for partitioning \hadi{a} CTU. The entire block is represented by a \textit{root node}\hadi{,} which has a depth value \hadi{of} $0$. Each node in the tree can have either four child nodes or zero child-node (\ie \textit{leaf node}). \hadi{The} depth level of the nodes increases by $1$ when traversed towards the bottom of the tree. If we consider the block size in \hadi{the} root node as $l_{max}\times l_{max}$ and \hadi{the} depth as $0$, then each sub-block at depth $d$ has the size $(l_{max}\times l_{max})/2^d$.}

In HEVC, CTUs have \hadi{a} predetermined size of up to $64\times64$ pixels\hadi{,} where $l_{max}= 64$. Each CTU can then be split recursively into square sized CUs. Each division increases the depth by $1$, \eg $64\times64$ is \textit{depth $0$} and $8\times8$ is \textit{depth $3$}. An example of a CTU quad-tree is shown in Fig.~\ref{fig:ctu}. The difference in block partitioning between AVC and HEVC is shown in Fig.~\ref{fig:avchevc}. \ekrem{We can clearly see the effect of different maximum block sizes between two codecs ($16\times16$ \hadi{for} AVC and $64\times64$ \hadi{for HEVC}). Both codecs use smaller block sizes, \ie greater depths, for areas containing more \hadi{texture or motion complexity}  \eg center of the frame\hadi{,} where the players move. 
\hadi{It can also be seen that HEVC takes larger CUs for the areas with less texture or motion information, which results in a saving of more bitrates.}}   

Furthermore, coding unit (CU), prediction unit (PU), and transform unit (TU) concepts have been introduced in HEVC which, are associated with a CTU. 
 
\subsection{Coding Unit (CU)}
\begin{figure*}[pos=t]
\captionsetup{justification=centering,
              singlelinecheck=false,
              font=sf, labelfont=bf} 
\centering
\includegraphics[width=0.9\textwidth]{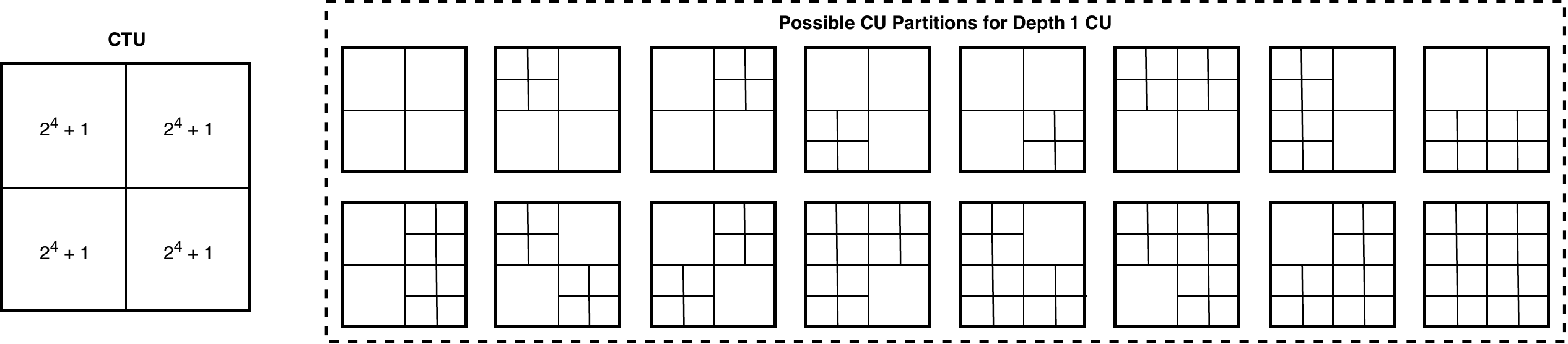}
\caption{Illustration of possible CU partitions.}
\label{fig:possiblecus}
\end{figure*}

Each leaf node of a quad-tree, representing a \hadi{square} region inside the CTU is called CU that can be from $8\times8$ to $64\times64$ \hadi{pixels}. Fig.~\ref{fig:possiblecus} exemplifies a CTU partitioning that contains 16 CUs or leaf nodes with sizes \hadi{from} $8\times8$ to $32\times32$ \hadi{pixels}. For each CU, three Coding Blocks (CBs) are associated in the video frame buffer, one for luma ($Y$) sample, and two for chroma ($Cb$, $Cr$) samples.

\subsection{Prediction Unit (PU)}
 
\begin{figure}[pos=t]
\captionsetup{justification=centering,
              singlelinecheck=false,
              font=sf, labelfont=bf} 
\centering
\includegraphics[width=0.45\textwidth]{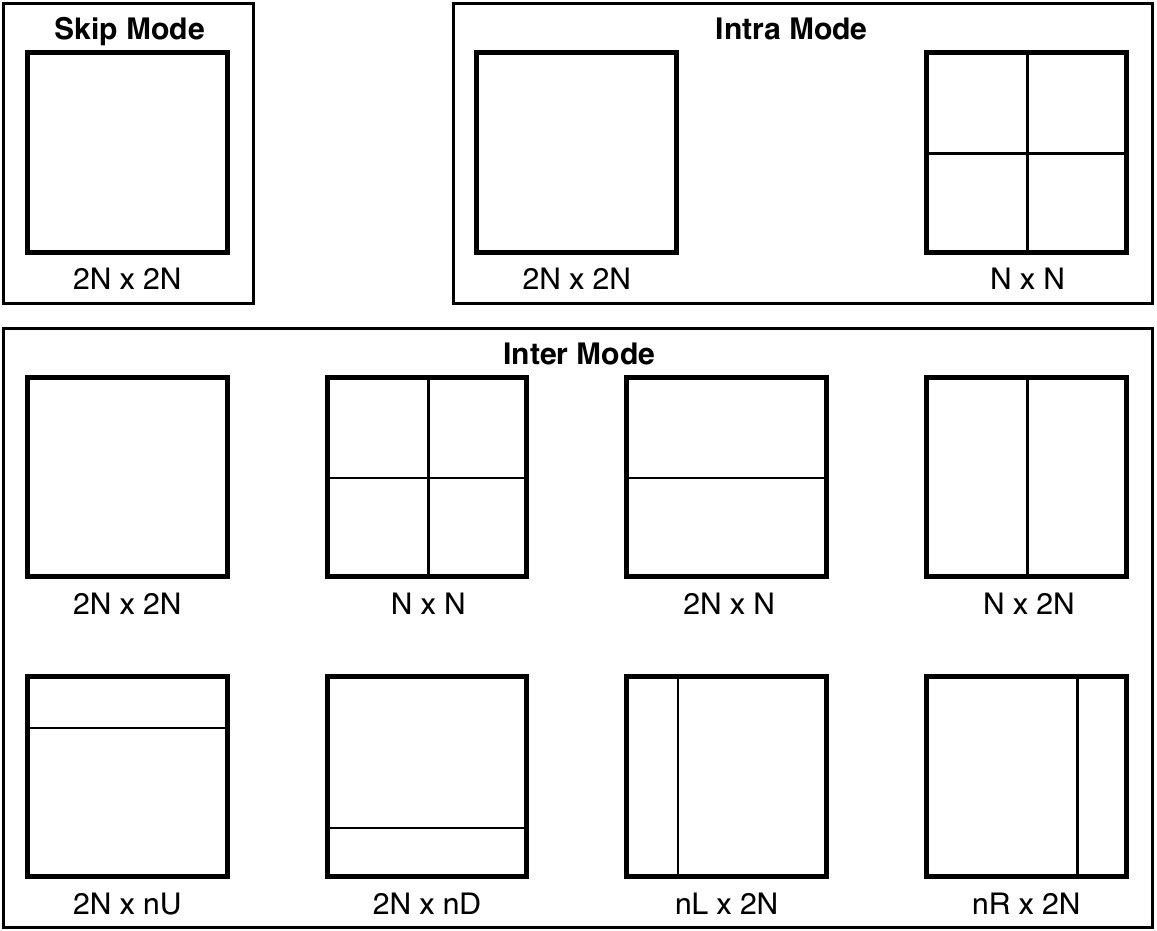}
\caption{PU splitting modes.}
\label{fig:pu}
\end{figure}

A decision on whether to perform inter- or intra-picture prediction on a block is made at the CU level. Each CU is split into PUs according to \hadi{the} prediction type. As with AVC, three prediction types are available for each CU: \textit{(i)} inter-coded CU, \textit{(ii)} skipped CU, and \textit{(iii)} intra coded CU. Various PU modes are illustrated in Fig.~\ref{fig:pu}.

\subsubsection{Inter Coded CUs}
There are eight different modes for \hadi{the} inter-picture prediction type, \ie four square- or rectangular-shaped and four asymmetric modes. A CU with \hadi{a} size \hadi{of} $2N\times2N$ can be split into one of the following modes: single PU with size ($PART\_2N\times2N$), four PUs with sizes ($PART\_N\times N$), two PUs with sizes ($PART\_2N\times N$) or ($PART\_N\times 2N$) for square- or rectangular-shaped modes. A CU can \hadi{be} further split into two PUs with sizes ($PART\_2N\times nU$), ($PART\_2N\times nD$), ($PART\_nL\times 2N$) or ($PART\_nD\times 2N$) in the asymmetric mode.  
It should be noted that ($PART\_N\times N$) is checked only when a CU has its minimum size ($2N\times 2N = 8\times 8$ for main profile). For other CU sizes, ($PART\_N\times N$) is similar to splitting the CU into four smaller sub-CUs. Also, asymmetric modes are disabled when CU size is equal to $8\times 8$ to reduce the complexity. 
 
\subsubsection{Skipped CUs}
Skipped CU mode is a special inter-coded CU mode where both motion vector and residual energy are zero. For each $2N\times2N$ CU, only one $2N\times2N$ PU is considered for the skipped mode.   

\subsubsection{Intra Coded CUs}
Two PU modes are available for intra-picture coding of a CU, ($PART\_2N\times 2N$) and ($PART\_N\times N$). Similar to CU, three prediction blocks (PBs) are considered in the video frame buffer for each color component. 

\subsection{Transform Unit (TU)}
When the optimal prediction mode is selected for each leaf CU, residual errors are transformed into \hadi{a} TU, and a residual quad-tree (RQT) structure is used to determine the best TU partitioning for each leaf CU. Square-shaped TUs with sizes between $4\times 4$ to $32\times 32$ \hadi{samples} are supported in HEVC~\cite{TU}.

\subsection{\ekrem{Overview of HEVC CTU Partitioning}}
{There are numerous possible patterns for the division of a single CTU. It is easy to understand that from a bottom-up approach. Let us assume a CU with a depth $2$. For this particular CU, there are two options, split or non-split. Now\hadi{,} if we go up one level to depth $1$, we have four possible sub-CUs with depth $2$ and two possible partitions for each sub-CU. Thus, for depth $1$, there are $2^4$ possible sub-CU partitions and also one more option that is non-split, which gives a total of $2^4 + 1$ options in total for a single depth $1$ CU. Following the same approach, if we go up one more level to depth $0$, there will be $(2^4 +1)^4$ possible partitions for depth $0$. Again, we also have the option to not split, so there is a total of $(2^4 + 1)^4 + 1 = 83522$ possible partitions for a single CTU. 
To find an optimal CTU partitioning from the $83522$ possible partitions, HEVC searches 85 CUs with different sizes for each CTU. These $85$ CUs are: one $64\times 64$, four $32\times 32$, sixteen $16\times 16$, and sixty four $8\times 8$ pixels blocks.

In addition to finding the correct CU depth structure, the PU modes and the TU partitioning for each possible CU must also be correctly determined. Thus, the search for the optimal CTU structure using a brute force approach to determine the one with the minimum rate-distortion (RD) cost using a Lagrangian multiplier, takes the largest amount of time in the encoding process\revision{~\cite{quadtree3}.} To show how eliminating one CTU depth search affects encoding efficiency and time-complexity, we encoded several sequences where the depth $0$ is eliminated by setting the maximum CTU size to $32 \times 32$ pixels. Moreover, we also eliminated depth $3$ by setting \hadi{the} minimum CU size to $16\times 16$ \hadi{pixels}. \hadi{The} results have been summarized in Table~\ref{tab:depthstudy}. \ekrem{In these tables, \hadi{the} classes represent the category of videos based on the video resolution~\cite{HEVC_CTC}. \textit{All Intra}, \textit{Random Access}, and \textit{Low Delay B} are the HEVC configuration files used during the encoding~\cite{HEVC_CTC}. All reported results are obtained by comparing the encoding results with HEVC reference software (HM 16.20) using a modified configure\hadi{ation}, which limits the minimum or maximum CU sizes, and the unmodified configure.} It is clear that limiting the depth of CTU partitioning can reduce the time complexity at the cost of bitrate increase.} 

\begin{table*}[pos=t]
\caption{\revision{Encoding time ($\Delta_{E}$), decoding time ($\Delta_{D}$), BD-PSNR (BDP), and BD-Rate (BDR) when depth 0 (No D0) or depth 3 (No D3) is skipped.}}
\label{tab:depthstudy}
\resizebox{\textwidth}{!}{%
\begin{tabular}{c|cccc|cccc|cccc|cccc|cccc|cccc}
\multirow{3}{*}{\textbf{Class}} & \multicolumn{8}{c|}{\textbf{All Intra}}                                                                                                                                                                                                                                                         & \multicolumn{8}{c|}{\textbf{Random Access}}                                                                                                                                                                                                                                                     & \multicolumn{8}{c}{\textbf{Low Delay B}}                                                                                                                                                                                                                                                       \\ \cline{2-25} 
                                & \multicolumn{4}{c|}{\textbf{No D0}}                                                                                                            & \multicolumn{4}{c|}{\textbf{No D3}}                                                                                                            & \multicolumn{4}{c|}{\textbf{No D0}}                                                                                                            & \multicolumn{4}{c|}{\textbf{No D3}}                                                                                                            & \multicolumn{4}{c|}{\textbf{No D0}}                                                                                                            & \multicolumn{4}{c}{\textbf{No D3}}                                                                                                            \\ \cline{2-25} 
                                & \multicolumn{1}{c|}{\textbf{$\Delta_{E}$}} & \multicolumn{1}{c|}{\textbf{$\Delta_{D}$}} & \multicolumn{1}{c|}{\textbf{BDP}} & \textbf{BDR}     & \multicolumn{1}{c|}{\textbf{$\Delta_{E}$}} & \multicolumn{1}{c|}{\textbf{$\Delta_{D}$}} & \multicolumn{1}{c|}{\textbf{BDP}} & \textbf{BDR}     & \multicolumn{1}{c|}{\textbf{$\Delta_{E}$}} & \multicolumn{1}{c|}{\textbf{$\Delta_{D}$}} & \multicolumn{1}{c|}{\textbf{BDP}} & \textbf{BDR}     & \multicolumn{1}{c|}{\textbf{$\Delta_{E}$}} & \multicolumn{1}{c|}{\textbf{$\Delta_{D}$}} & \multicolumn{1}{c|}{\textbf{BDP}} & \textbf{BDR}     & \multicolumn{1}{c|}{\textbf{$\Delta_{E}$}} & \multicolumn{1}{c|}{\textbf{$\Delta_{D}$}} & \multicolumn{1}{c|}{\textbf{BDP}} & \textbf{BDR}     & \multicolumn{1}{c|}{\textbf{$\Delta_{E}$}} & \multicolumn{1}{c|}{\textbf{$\Delta_{D}$}} & \multicolumn{1}{c|}{\textbf{BDP}} & \textbf{BDR}     \\ \hline
Class A                         & 88 \%                                      & 103 \%                                     & -0.40 \%                          & 0.81 \%          & 58 \%                                      & 94 \%                                      & -1.02 \%                          & 2.11 \%          & 84 \%                                      & 118 \%                                     & -0.73 \%                          & 1.91 \%          & 72 \%                                      & 98 \%                                      & -1.36 \%                          & 3.64 \%          & 83 \%                                      & 113 \%                                     & -0.72 \%                          & 2.09 \%          & 72 \%                                      & 96 \%                                      & -1.13 \%                          & 3.30 \%          \\
Class B                         & 89 \%                                      & 110 \%                                     & -0.16 \%                          & 0.48 \%          & 57 \%                                      & 99 \%                                      & -0.03 \%                          & 0.09 \%          & 86 \%                                      & 113 \%                                     & -0.67 \%                          & 2.93 \%          & 72 \%                                      & 99 \%                                      & -0.03 \%                          & 0.09 \%          & 85 \%                                      & 112 \%                                     & -0.45 \%                          & 1.83 \%          & 73 \%                                      & 99 \%                                      & -0.03 \%                          & 0.02 \%          \\
Class C                         & 90 \%                                      & 106 \%                                     & -0.20 \%                          & 0.87 \%          & 59 \%                                      & 97 \%                                      & -0.35 \%                          & 1.43 \%          & 86 \%                                      & 115 \%                                     & -0.68 \%                          & 2.26 \%          & 73 \%                                      & 98 \%                                      & -0.29 \%                          & 1.40 \%          & 85 \%                                      & 113 \%                                     & -0.58 \%                          & 2.71 \%          & 75 \%                                      & 99 \%                                      & -0.26 \%                          & 1.22 \%          \\
Class D                         & 89 \%                                      & 101 \%                                     & -0.26 \%                          & 0.39 \%          & 61 \%                                      & 96 \%                                      & -5.78 \%                          & 6.74 \%          & 87 \%                                      & 103 \%                                     & -0.05 \%                          & 0.25 \%          & 71 \%                                      & 97 \%                                      & -4.41 \%                          & 8.77 \%          & 88 \%                                      & 103 \%                                     & -0.29 \%                          & 0.61 \%          & 71 \%                                      & 95 \%                                      & -4.03 \%                          & 8.91 \%          \\
Class E                         & 89 \%                                      & 102 \%                                     & -0.58 \%                          & 1.46 \%          & 57 \%                                      & 97 \%                                      & -1.32 \%                          & 3.33 \%          & 86 \%                                      & 124 \%                                     & -1.13 \%                          & 3.87 \%          & 72 \%                                      & 99 \%                                      & -1.09 \%                          & 3.88 \%          & 85 \%                                      & 116 \%                                     & -1.30 \%                          & 4.79 \%          & 73 \%                                      & 99 \%                                      & -1.00 \%                          & 3.83 \%          \\ \hline
\textbf{Average}                & \textbf{89 \%}                             & \textbf{104 \%}                            & \textbf{-0.32 \%}                 & \textbf{0.79 \%} & \textbf{58 \%}                             & \textbf{96 \%}                             & \textbf{-2.02 \%}                 & \textbf{3.25 \%} & \textbf{86 \%}                             & \textbf{114 \%}                            & \textbf{-0.65 \%}                 & \textbf{2.16 \%} & \textbf{72 \%}                             & \textbf{98 \%}                             & \textbf{-1.69 \%}                 & \textbf{3.92 \%} & \textbf{85 \%}                             & \textbf{111 \%}                            & \textbf{-0.68 \%}                 & \textbf{2.42 \%} & \textbf{73 \%}                             & \textbf{97 \%}                             & \textbf{-1.49 \%}                 & \textbf{3.81 \%}
\end{tabular}}
\end{table*}

\begin{figure*}[pos=t]
    \captionsetup{justification=raggedright,
                  singlelinecheck=false,
                  font=sf, labelfont=bf} 
    \centering
    \includegraphics[width=0.9\textwidth]{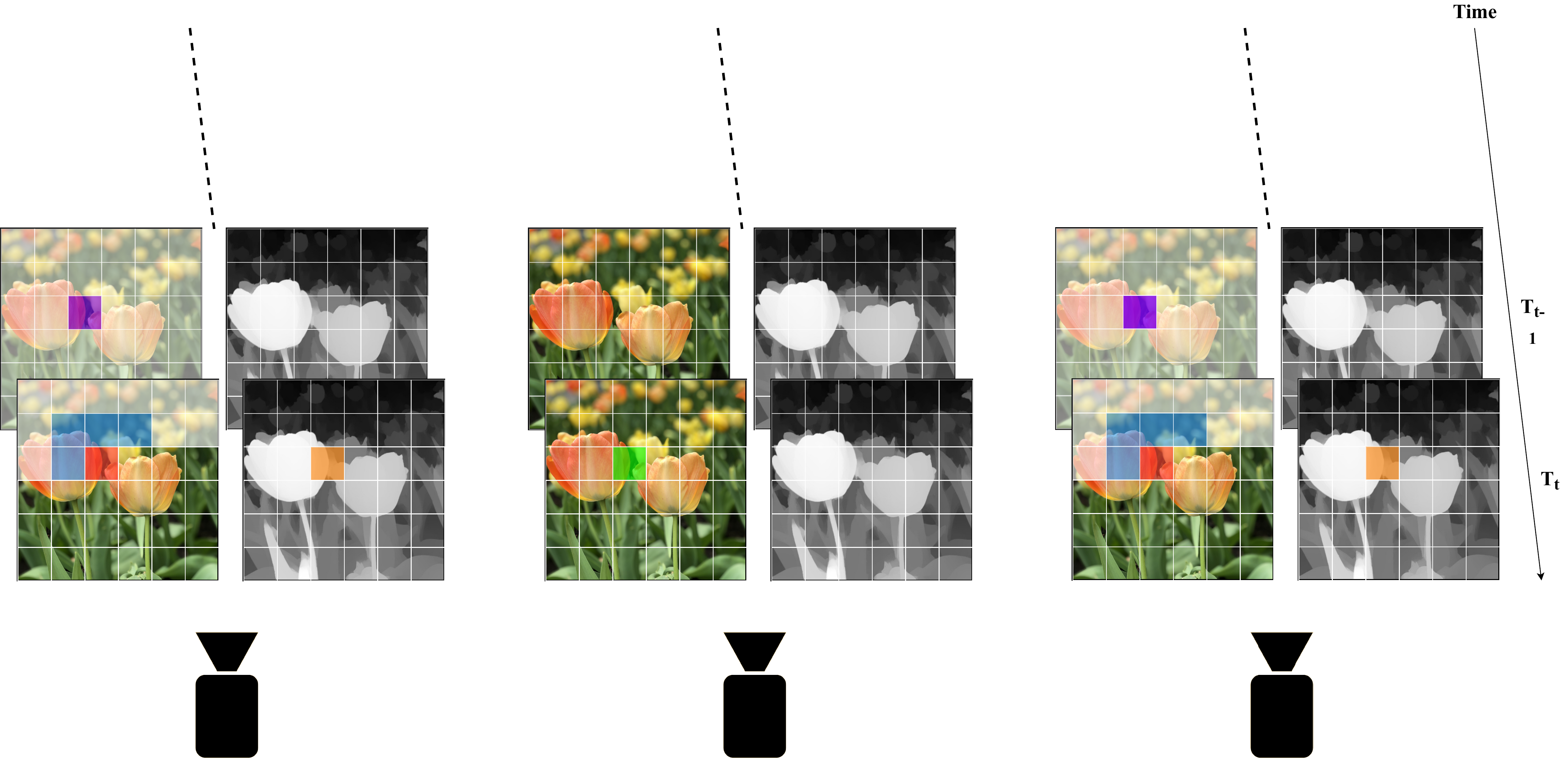}
    \caption{CTUs are shown in different colors. \textit{\textbf{Red}}: Current CTU, \textit{\textbf{Blue}}: Four spatially neighboring CTUs, \textit{\textbf{Purple}}: Temporally neighboring CTU, \textit{\textbf{Green}}: Co-located interview CTU in the base view for MV-HEVC and 3D-HEVC, and \textit{\textbf{Orange}}: Co-located CTU in the depth frame for 3D-HEVC.}
    \label{fig:3dhevc}
\end{figure*}

\begin{figure*}[pos=t]
    \captionsetup{justification=raggedright,
                  singlelinecheck=false,
                  font=sf, labelfont=bf} 
    \centering
    \includegraphics[width=0.9\textwidth]{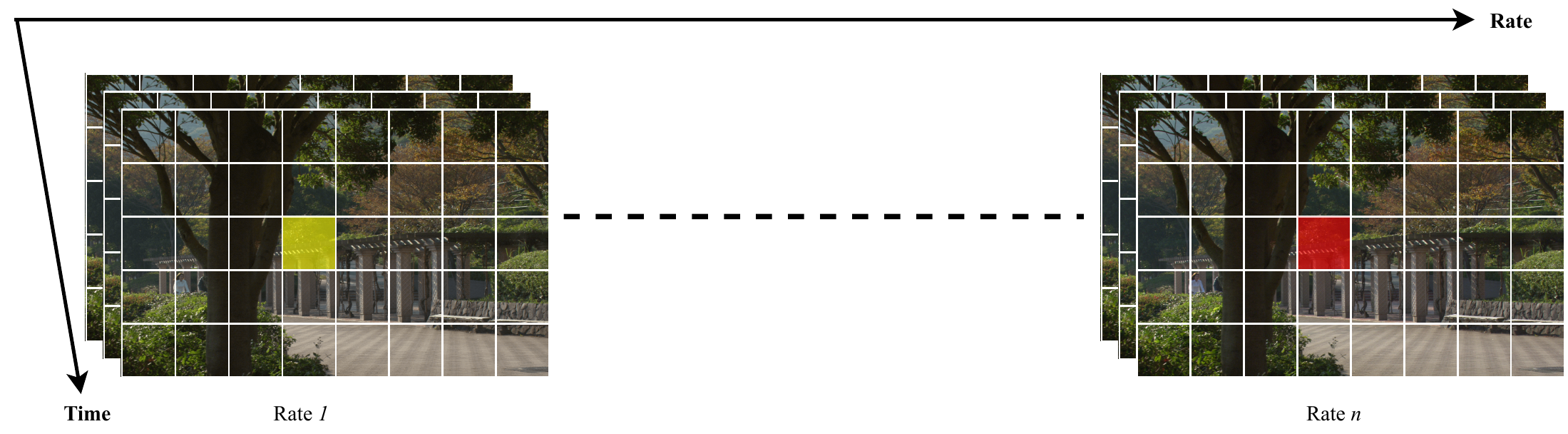}
    \caption{Multirate encoding. Red block represents current CTU while yellow block represents co-located CTU in the previously encoded representation.}
    \label{fig:multirate}
\end{figure*}

{To overcome the high complexity raised from \hadi{the} CTU partitioning in HEVC, many algorithms have been proposed which try to use \hadi{the} similarity between \hadi{the} partitioning of spatial and/or temporal neighboring CTUs or inherent features of each CTU, to skip \hadi{the search for} unnecessary CUs. For each non-border CTU, there are four spatially neighboring CTUs in the same frame and one temporally co-located CTU in the reference frame. \ekrem{Moreover, additional neighboring CTUs can be found in the HEVC extensions, which are designed for specific scenarios. We briefly introduce them in the following section.}}

\ct{\subsection{Overview of HEVC Extensions}}
\revision{Multiple extensions to HEVC are proposed to address challenges in various use cases~\cite{HEVC_extensions}. These extensions have been specified in the following order:
\begin{enumerate}
    \item HEVC Format Range Extension (REXt)~\cite{HEVC_RExt}
    \item Scalable HEVC (SHVC)~\cite{SHVC}
    \item Multiview High Efficiency Video Coding (MV-HEVC)~\cite{MV-HEVC}
    \item 3D High Efficiency Video Coding (3D-HEVC)~\cite{MV-HEVC}
    \item HEVC Screen Content Coding Extension (SCC)~\cite{HEVCSCC}
\end{enumerate}
}

\revision{The Range Extension of HEVC (RExt)~\cite{HEVC_RExt} provides support for  high bit depths beyond 10 bits per sample, and various  chroma sampling formats including monochrome, 4:2:2, and 4:4:4.}

\revision{Scalable HEVC (SHVC)~\cite{SHVC} provides support for spatial, signal-to-noise ratio, bit-depth, and color gamut scalability in addition to the temporal scalability that was supported by the first version of HEVC. In scalable video coding, a video is encoded in multiple layers. The lowest quality representation, referred to as the base layer (BL), is first encoded. It is used as a reference to encode one or more enhancement layers (ELs) with improved video quality in terms of various scalability dimensions. In addition to the available spatial and temporal CTUs in HEVC, interlayer CTUs can be used for prediction. From different scalabilities, scalable-quality or scalable-spatiality are typically used for CTU prediction. }

{MV-HEVC~\cite{MV-HEVC} is an extension of HEVC, which allows efficient encoding of multiple camera views by enabling the use of interview references in motion-compensated prediction. Views are divided into base and dependent views. Base views are encoded using HEVC simulcast\hadi{,} \ekrem{\ie each view is encoded independently}.

Dependent views exploit dependencies between the views and use reconstructed base view frames as additional reference frames for motion compensation. Therefore, in addition to the available spatial and temporal CTUs in HEVC, interview CTU, \ie co-located CTU in the base view, can be used for prediction. The interview co-located CTU is illustrated in Fig.~\ref{fig:3dhevc}.} %

{The 3D extension of HEVC (3D-HEVC)~\cite{MV-HEVC} is an extension of HEVC that supports the encoding of multiple views and their associated depth information. Similar to MV-HEVC, texture frames can utilize information from four spatially neighboring CTUs, temporally co-located CTU(s), and interview CTU(s). In addition to the above-mentioned CTUs, information of co-located CTU in the corresponding depth frame can also be used if \hadi{the} depth frame has been previously encoded. Otherwise, \hadi{the} co-located CTU in the associated texture frame can be used to predict \hadi{the} depth level of the depth frame. Fig.~\ref{fig:3dhevc} shows the association of the texture and depth frames.}%

{HEVC Screen Content Coding (SCC)~\cite{HEVCSCC} has been developed to provide improved \hadi{en}coding efficiency for videos that contain \hadi{a} significant amount of screen-captured content, \hadi{as} characteristics of these videos differ from those of camera-captured contents. To achieve this, several tools have been added to the basic HEVC that are specifically designed for screen content. One \hadi{of these tools} is Intra-Block Copy (IBC), which is another CU mode that is added along with \hadi{the} existing conventional Intra (Cintra) and Inter modes. It can be considered as motion estimation inside a frame at the PU level. When a CU is encoded in IBC, PUs of this CU are searched for similar reconstructed blocks within the current frame. Another tool is Palette Mode \ekrem{(PLT)}, which focuses on color information since screen content videos usually contain \hadi{a} a small number of different colors. PLT first enumerates each distinct color in the block, and these indexes are used for each sample rather than actual values to define color information. 
Also, Adaptive Color Transform (ACT) is proposed for color-coding since the majority of screen content videos use \hadi{the} RGB color space and not YCbCr. 
In HEVC-SCC, an image block can be encoded directly in the RGB space or can be converted to \hadi{the} YCoCg space during encoding\hadi{,} depending on the content characteristic of the block. Finally, Adaptive Motion Vector Resolution (AMVR) is added to deal with discrete motion in the screen content video\hadi{,} which can be represented by integer motion vectors. This is because, in screen-captured videos, movement is precisely aligned with pixels in general. AMVR allows the motion vectors to switch between integer and fractional values.}

\ekrem{\subsection{Multirate Encoding}}
{ Adaptive HTTP streaming~\cite{DASH:2011,DASH_IEEE} provides multiple representations of the same content in different qualities and resolutions. This allows clients to request segments in a dynamic and adaptive way depending on the network conditions. When a representation is encoded, \hadi{the} CTU depth information \hadi{from that representation} can be used by the other representations. Therefore, in addition to the four spatially and one temporally co-located CTUs, a co-located CTU in the other representation can also be used to increase the accuracy of \hadi{the} depth prediction for the current CTU. Fig.~\ref{fig:multirate} shows co-located CTU in the previously encoded representation.

\begin{figure}[pos=t]
    \captionsetup{justification=centering,
                  singlelinecheck=false,
                  font=sf, labelfont=bf} 
    \centering
    \includegraphics[width=\linewidth]{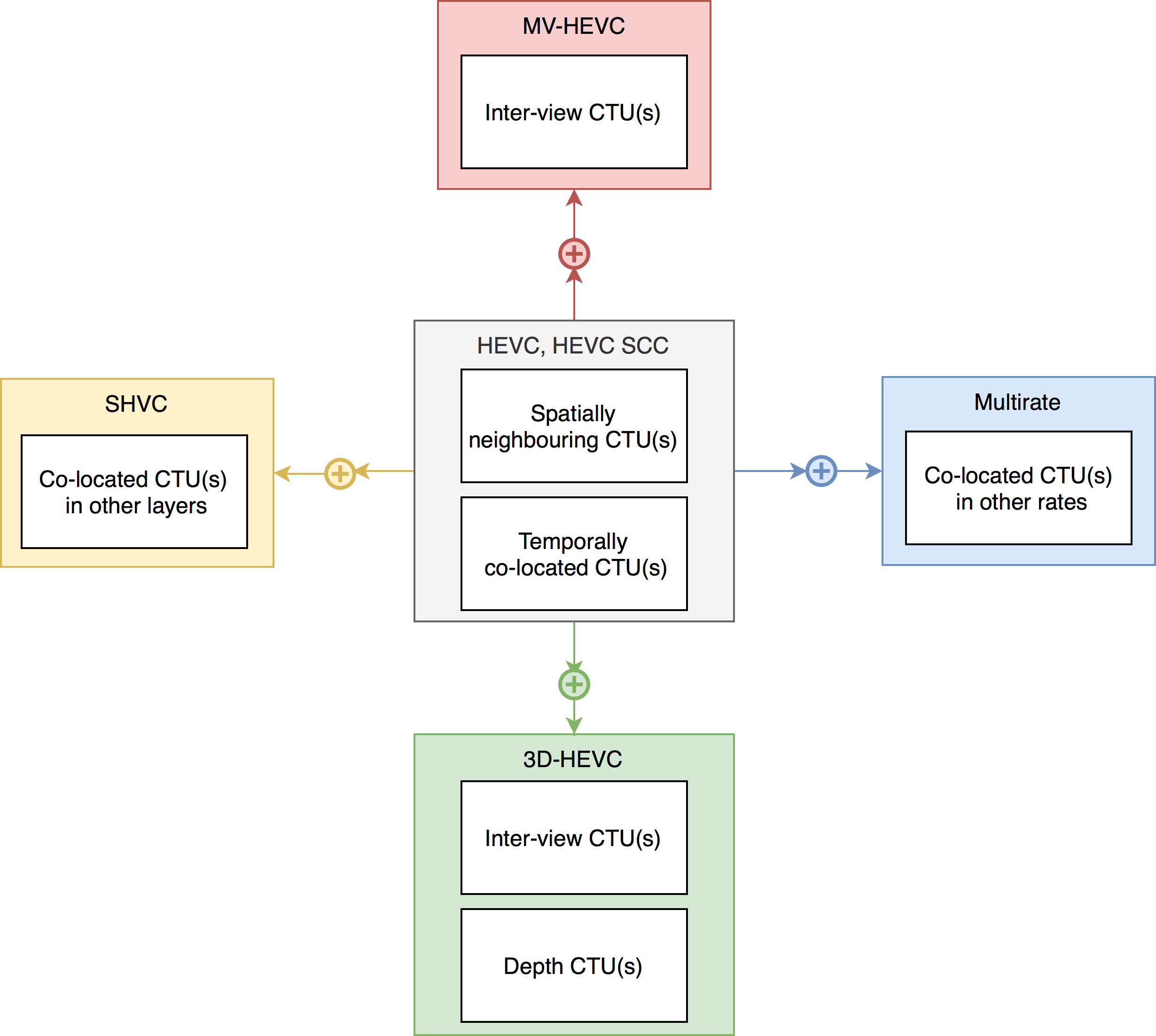}
    \caption{Available CTUs for HEVC and its extensions}
    \label{fig:NeighboringCTUs}
\end{figure}
}

\ekrem{\subsection{Summary}}
Fig.~\ref{fig:NeighboringCTUs} summarizes the available CTUs that can be used by the current CTU for different HEVC extensions. \ekrem{Four spatially neighboring CTUs and \hadi{one} co-located CTU in the temporally neighboring frame are available for HEVC and all its extensions. Interview CTU is added to the list of available reference CTUs for MV-HEVC\hadi{,} and \hadi{the} depth CTU is also available for 3D-HEVC together with all the CTUs mentioned above. Finally, co-located CTUs from previously encoded representations are also available for multirate encoding \revision{while co-located CTUs in different layers are available for SHVC.}} 

Since complex CTUs have \hadi{a} larger depth than homogeneous CTUs, some features are extracted to measure the complexity of \hadi{individual} CTUs and then \hadi{to} make a decision on their depth levels. 
\ekrem{Numerous approaches have been proposed to improve CTU partitioning decision of HEVC to eliminate unnecessary searches and improve overall encoding pipeline. In the following section, we provide an overview of existing statistics based approaches in the literature.}

\section{Statistics Based Approaches}
\label{sec:statistic}

Statistics based approaches exploit \hadi{the} statistical similarity in the video to reduce the time needed for the CU depth search. 
\hadi{Since video is composed of successive frames, these approaches can achieve a significant reduction in time. These methods exploit previously known correlations in the video by manually defining \hadi{a} set of features, thresholds, rules\ct{, \etc}}    \ct{The m}ain advantage of these approaches \hadi{is that} they are faster and less complex \hadi{than approaches based on machine learning, with the disadvantage of being less general and less improvement in the encoding efficiency.}

In this section, statistics based approaches are categorized into \ct{\textit{two groups}} based on the main feature used by the \ct{respective} approach: \ct{\textit{(i)} \textit{Neighboring}} CTUs use information that is available from spatially and temporally neighboring CTUs and \ct{\textit{(ii)} \textit{inherent}} approaches only use information that is available in the current CTU.

\ekrem{\subsection{Neighboring CTUs Approaches}}

\ct{The \textit{neighboring CTUs} approaches can be further divided into approaches defined as part of the \textit{(a)} core \textit{HEVC} specification, \textit{(b)} \textit{HEVC extensions}, and \textit{(c)} \textit{multirate} HEVC techniques.}

\subsubsection{Core \textit{HEVC} specification}
There are several methods in the literature that exploit \hadi{only} temporally or spatially closed CU information. 

\revision{Kim~\etal~\cite{H_NeighbourCU_2} propose to determine a search range for the current CTU using a threshold value estimated from the spatially neighboring CTUs which is calculated as Eq.~\ref{eq:1}.
\begin{equation}
    \theta = \sum_{i=0}^{3} D_i
\label{eq:1}
\end{equation}
where $\theta$ is the threshold value and $D_{i}$ is the depth value of the neighboring CTU. Corresponding CU depth ranges for the calculated threshold values are listed in Table \ref{table:neighbourcu_2}.
}%

Similar to~\cite{H_NeighbourCU_2},~Shen~\etal~\cite{H_ContentFastCU} use information from four spatially neighboring CTUs, but to categorize CTUs into four texture classes ranging from homogeneous to rich texture, using the formula in Eq.~\ref{eq:neighbourcu2}.

\revision{
\begin{equation}
\label{eq:neighbourcu2}
    D_{pre} = \sum_{i=0}^{3}  \alpha_i D_i
\end{equation}
where $D_{i}$ is the depth value and $\alpha_i$ is a weighting factor assigned to each neighboring CTU. Based on $D_{pre}$, texture type of the current CTU is determined as given in Table \ref{table:contentfastcu}, and based on the $D_{pre}$, maximum or minimum depth level is changed for the current CTU.}

\revision{Hsu~\etal~\cite{H_NeighbourCU}} propose three criteria for the CU split decision, \ie \textit{(i)} co-located CTU in the previous frame has larger depth, \textit{(ii)} all neighboring CTUs have larger depths, and \textit{(iii)} the current frame is not an I-frame. On top of that, another three criteria are used for early termination, \ie \textit{(i)} co-located CTU has smaller depth, \textit{(ii)} at least three neighbours have smaller depths, and \textit{(iii)} the current frame is not an I-frame. Under the other conditions, the unmodified HEVC search is used.%

\begin{table}[pos=t]
\centering
\caption{\revision{Threshold values and their corresponding depth ranges used in \cite{H_NeighbourCU_2}.}}
\label{table:neighbourcu_2}
\begin{tabular}{c|c}
\Xhline{2\arrayrulewidth}
Threshold Value     & Current CTU depth range \\ \hline
$\theta = 0$ & 0 to 1                  \\
$1\leqslant \theta \leqslant 3$ & 0 to 2  \\
$4\leqslant \theta \leqslant 6$ & 0 to 2   \\
$7\leqslant \theta \leqslant 12$ & 1 to 3   \\
\Xhline{2\arrayrulewidth}
\end{tabular}
\end{table}
\begin{table}[pos=t]
\centering
\caption{\revision{CTU texture categorization criteria used in \cite{H_ContentFastCU}.}}
\label{table:contentfastcu}
\begin{tabular}{c|c}
\Xhline{2\arrayrulewidth}
 $D_{pre}$  & Current CTU texture type \\ \hline
$0.5 \leqslant D_{pre}$  & type I  \\
$0.5 < D_{pre}\leqslant 1.5$  & type II \\
$0.5 < D_{pre}\leqslant 2.5$  & type III \\
$ D_{pre} > 2.5$  & type IV  \\
\Xhline{2\arrayrulewidth}
\end{tabular}
\end{table}

{Another approach \ekrem{proposed by Cen~\etal~\cite{H_SeqSim}} follows \hadi{the} average depth value of the four spatially neighboring CTUs. This average value is then compared to a predefined threshold value to determine \hadi{if the} depth range of \hadi{the} current CTU is $\{0~2\}$ or $\{1~3\}$.}%

{\ekrem{Li~\etal~\cite{H_Combined} use spatially neighboring CTUs and an RD cost threshold together.} \hadi{The} RD cost threshold is updated online using training frames. Left and upper CTUs are used, and \hadi{the} depth search range for \hadi{the} current CU is obtained by limiting it between the minimum and maximum depths found in these two CTUs. If the  \hadi{depth of the} current CU is smaller than $depth_{min}$, it is split, and if it is equal to or larger than $depth_{max}$\hadi{,} CU is not split. Also, an early termination \hadi{algorithm} is proposed based on RD cost as an alternative method for all intra configuration only, \ekrem{which exploits the temporal correlation between adjacent frames.} For this approach, frames are periodically categorized into training and test frames. \hadi{The} first frame of each period is \hadi{a} training frame which is used to extract statistical information about CU depth levels and RD costs. This information is \ekrem{then} used to determine a threshold for early termination in test frames. These two approaches are combined for all intra configuration.}%

{A two-level depth decision algorithm is proposed based on previously encoded frames  \ekrem{by Huade~\etal~\cite{H_CULimitFrame}}. At the frame level, it is supposed that the depth level of all CUs within a frame are concentrated on two depths. These two depth levels are the most common CU depth levels found in the latest encoded frame of the same type.  In CU level, \hadi{the} depth of each CU is limited to the minimum and maximum depth values found in the temporally co-located CU.} %

{\ekrem{Li~\etal~\cite{H_InterCUEn} use \hadi{the} maximum depth of \hadi{the} temporally co-located CU ($D_{pre0}$) to predict \hadi{the} depth of \hadi{the} current CU ($D_{cur}$) using Eq.~\ref{eq:intercuen_1}.}
    \begin{equation}
    \label{eq:intercuen_1}
     D_{cur} = D_{pre0} + \Delta D
    \end{equation}
where $\Delta D$ is variation between $D_{cur}$ and $D_{pre0}$ which is determined based on \hadi{encoding} parameters \ekrem{as shown in Eq.~\ref{eq:intercuen_2}.}
    \begin{equation}
    \label{eq:intercuen_2}
     \Delta D = |D_{pre0} - D_{pre1}| + |QP_{cur} - QP_{pre0}| - |QP_{pre0} - QP_{pre1}|
    \end{equation}
}

{\ekrem{Amirpour~\etal~\cite{LF1,LF2} use co-located CTUs to limit \hadi{the} depth search range for the current CTU.} Four co-located CTUs in four reference frames are replaced with four spatially neighboring CTUs to determine \hadi{the} depth range for the current CTU. Their minimum and maximum depth values are used to limit \hadi{the} depth range for the current CTU.}%

{\ekrem{Pan~\etal~\cite{H_FrameAndPU} use} co-located CTU in addition to spatially neighboring CTUs to make a decision on the depth level of the current CTU. CTU partitioning is terminated in depth 0 if the following condition \ekrem{in Eq.~\ref{eq:frameandpu}} is satisfied:
\begin{equation}
\label{eq:frameandpu}
\centering
 W = \sum_{n=0}^{4} \lambda_n C_n \geqslant \alpha
\end{equation}

where $\lambda_n$ is a weighting factor and $C_n$ represents the weight of the reference CTUs, which is $1$ if the reference CTU is encoded as depth $0$, otherwise, $C_n$ is equal to $0$; Depth $3$ searches are skipped based on \hadi{the} correlation between PU mode decisions.}%

{CTUs are classified into \textit{simple} and \textit{complex} CTUs in \ekrem{\hadi{the} method proposed by Zhou~\etal~\cite{H_Spatiotemporal}}. If \hadi{the} maximum depth of a CTU is 0 or 1\hadi{,} it is considered \hadi{as a} simple \hadi{CTU}, otherwise, complex. A depth decision algorithm has been proposed based on the complexity of the left, upper, and co-located CTUs. If all above-mentioned neighboring CTUs are complex, depth 0 is skipped from searching, and if they are all simple\hadi{,} depth 3 is skipped from the search process.}%

{\ekrem{Leng~\etal~\cite{H_CBFrameSkip} predict \hadi{the} depth of CTUs both at frame and CTU level\hadi{s}}. Rarely used depths in the reference frame are skipped in the current frame. In the CTU level, partitioning of \hadi{the} current CU is terminated if \textit{(i)} \hadi{the} current depth is equal to depth of \hadi{the} co-located CU, or \textit{(ii)} if \hadi{the} current depth is equal to \hadi{the} depth of two or more spatially neighboring CTUs.}%

{\ekrem{Shen~\etal~\cite{H_HEVCEff} propose to limit CU depth search range} using the information in four spatially neighboring CTUs and co-located CTU. Information from these sources is weighted according to its importance, CTUs in the horizontal and vertical direction\hadi{s} are given more weight, and \hadi{are} used twitfor depth prediction. Depending on the result of prediction, the current CU is classified into one of five types ranging from homogeneous to fast motion regions, and the depth is limited depending on the classification result.} %

{\ekrem{Correa~\etal~\cite{H_complexityControl} categorize frames into two groups}, \textit{unconstrained frames} ($F_u$) and \textit{constrained frames} ($F_c$). Unconstrained frames are encoded using the full RDO process and \hadi{the} remaining frames are encoded using \hadi{the} maximum depth of the co-located CTU in previous $F_u$ to limit CTU partitioning in the current CTU. \hadi{The} number of $F_c$ between $F_u$s is controlled by target complexity.}

{In another study, Correa~\etal~\cite{H_FrameEff_2} take into account the information of the spatially neighboring CTUs as well as the co-located CTU in the previous constrained frame $F_c$ on top of using the co-located CTU in the unconstrained frame $F_u$ to define maximum depth of the current CTU.}

{\ekrem{Bae~and~Sunwoo~\cite{H_DepthHistory} store CU depth information from CTUs in the previous five or six frames and use this information to decide early CU and PU terminations}. A weighted structure is used to give more importance to the closer frames. If all the depths are equal, then \hadi{the} same depth is selected as the final depth for the current CU and PU search is conducted. Otherwise, statistical properties of CTUs are calculated and used in the decision process.} %

{The depth distribution and RD cost of the co-located CTU are used to reduce \hadi{the} search range for the current CTU in \ekrem{method proposed by Park~\cite{H_CUDist}}. The search range is limited by the maximum and minimum CU depths that are used in the co-located CTU. To avoid error propagation in this approach, the search range is reset at every one-second interval, and those frames are referred to as reset frames for which the search range limitation is not applied.  A predefined threshold \hadi{value} is used for early termination, which is calculated based on the RD cost and CU depth distribution of co-located CTU.  If the RD cost of the current depth is higher than the threshold, the CU is further split. The threshold is adjusted based on the CU depth distribution in the co-located CTU in a way that if the co-located CTU contains more CUs at maximum depth, then the current CTU is less likely to be early terminated. Again, early termination is not applied for the reset frames.}%

\revision{Liu~\etal~\cite{H_RoughAccurate}} do partitioning of CTUs  in two steps, \ie \textit{rough} and \textit{accurate} determination. The rough determination step is used to predict the complexity of CU by using the depth values and prediction modes of neighboring CTUs. Instead of using information from the entire CTU, information in the edges of CTUs is used. For the left neighboring CTU, the rightmost $4\times64$ pixels area, and for the top neighboring CTU, the bottom-most $64\times4$ pixels areas are used. If the current CTU has a depth of $0$ and if there is a depth of $3$ in any of these areas, the RDO search is skipped, and the current CTU is split. Moreover, if the current CTU has a depth of $2$ and if there is a depth of $0$ in any of these areas, the CTU split will stop. In the accurate determination step, pixel values in the edges of the current CTU are used to make a decision. If the pixel values vary over a wide range, then this is accepted as an indicator of a complex CTU. On top of that, the entropy of pixel values and pixel differences in the top, left, right, and bottom edges are also used to calculate complexity. Each of these calculations is used to determine a threshold value for the complexity of the CTU. If the CTU is determined to be complex in any of these steps, it is split further; otherwise, CU splitting is stopped. The algorithm uses the decision of rough determination step unless the decision is uncertain. The accurate determination step is used only when the decision of the algorithm is uncertain.

{\ekrem{Zhao~\etal~\cite{H_sipc} propose a two-step depth decision algorithm.} First, a depth range $[D_{min}^{C} ~ D_{max}^{C}]$ is determined for the current CU (C) based on the depth values of left (L) and upper (U) CUs \ekrem{as in Eq.~\ref{eq:sipc}.}
\begin{equation}
\label{eq:sipc}
\begin{split}
    &D_{min}^{C} = min(D_{}^{L},D_{}^{U}) - 1 \\
    &D_{max}^{C} = max(D_{}^{L},D_{}^{U}) + 1 
\end{split}
\end{equation}

In the second step, if the RD costs of already searched for child CUs are larger than that of the parent CU, \hadi{the} search process is terminated.}%

\revision{\subsubsection{SHVC}}

\revision{Ge~\etal used the correlation between co-located CTUs in the base and quality enhancement layers to predict the depth of a CTU in the enhancement layer from the co-located CTU in the base layer~\cite{SHVC1}.
The depth of a CTU in the base layer is used as an upper bound for the co-located CTU in the quality enhancement layer. Therefore,  CUs with depth values higher than the maximum depth value of the co-located CTU in the base layer are skipped from searching in the enhancement layer.}

\revision{Wang~\etal~\cite{SHVC2} skip searching some depths for quality enhancement layer, based on the depth of the co-located CTU in the base layer as follows:}

\begin{itemize}
    \item If depth is 0 for a CTU in the base layer, search only depths 0 and 1 for the co-located CTU in the quality enhancement layer.
    
    \item If depth is 1 for a CTU in the base layer, search depths 0, 1, and 2 for the co-located CTU in the quality enhancement layer.
    
    \item If depth is 2 for a CTU in the base layer, search only depths 2 and 3 for the co-located CTU in the quality enhancement layer.
    
    \item If depth is 3 for a CTU in the base layer, search only depths 2 and 3 for the co-located CTU in the quality enhancement layer.
\end{itemize}

\revision{Li~\etal~\cite{SHVC3} use the co-located CTU in the base layer ($depth_{bl}$) in addition to the four co-located spatially neighboring CTUs in the quality enhancement layer to predict the depth value for the current CTU in the enhancement layer. It is predicted based on the following criteria:}
\begin{itemize}
    \item If $depth_{bl} = 0$ and sum of the depths for co-located spatially neighboring CTUs is less than 4, depths 2, and 3 are skipped.
    \item If $depth_{bl} \neq 0$ and sum of the depths for co-located spatially neighboring CTUs is greater than 10, depths 0, and 1 are skipped.
    \item Otherwise, all depths are searched. 
\end{itemize}

\revision{Wang~\etal~\cite{SHVC4} use two spatially neighboring CTUs, namely left and top CTUs from the enhancement layer, and the co-located CTU from the base layer to predict a CTU in quality enhancement layer. The RD cost is calculated for the same quad-tree structures of the three predictors, and the one with the minimum RD cost is selected as the quad-tree structure of the current CTU in the enhancement layer.}

\revision{Zuo~\etal~\cite{SHVC5} propose a CU depth early skip algorithm for the spatial enhancement layer as follows:}
\begin{itemize}
    \item If the depth of the corresponding CU in the base layer is equal to 0, CUs with depth 3 in intra mode, and CUs with depths 2, 3 in inter mode are skipped.
    
    \item If the depth of the corresponding CU in the base layer is equal to 1, CUs with depth 3 in both intra and inter modes are skipped.
    
    \item If the depth of the corresponding CU in the base layer is equal to 2, CUs with depth 1 in intra mode, and CUs with depths  3 in inter mode are skipped.
    
    \item If the depth of the corresponding CU in the base layer is equal to 3, CUs with depth 0 and 1 in intra mode are skipped.
    
\end{itemize}

Wali~\etal~\cite{SHVC6} encode the CU of the spatial enhanced layer as skip mode if its corresponding CU in the base layer was encoded in skip mode.

\revision{Lu~\etal~\cite{SHVC8} use the four spatially neighbouring CTUs and the corresponding CU in the base layer to determine the depth range for each CTU in  both quality and spatial enhancement layers.}

\revision{Dhollande~\etal~\cite{SHVC9} propose a fast UHD encoding method for SHVC. To reduce the encoding time-complexity of the UHD video, first the UHD version is obtained by using the SHVC multi-phase filter in the first phase, and then its encoding information is used to speed up encoding of the original UHD video. To encode the HD version, the maximum CU size is restricted to $32\times32$. Three different strategies were evaluated for CU sizes:}
\begin{itemize}
    \item S1: Search only the upsampled CU of a HD video in the UHD video. For example, if $16 \times16$ has been chosen for a CU in HD video, search only $32 \times32$ CU in the UHD.
    
    \item S2: Search the upsampled CU of a HD video and its four children in the UHD video. For example, if $16 \times16$ has been chosen for a CU in HD video, search  $32 \times32$ CU and its four $16 \times16$ children CUs in the UHD.
    
    \item S3: Search the upsampled CU of a HD video and all of its children in the UHD video. For example, if $16 \times16$ has been chosen for a CU in HD video, search  $32 \times32$ CU and its four $16 \times16$ children CUs, their sixteen  $8 \times8$ children CUs in the UHD.

\end{itemize}

\subsubsection{MV-HEVC and 3D-HEVC}

{\ekrem{Chi~\etal~\cite{MV-HEVC3} use} the maximum depth of the co-located CU in the base view as a threshold to limit splitting the current CU in the dependent views.}%

{\ekrem{Khan~and~Khattak~\cite{MV-HEVC1} use HEVC simulcast, \ie view\hadi{s} are encoded independently with HEVC, for base views.} 
For dependent views, \hadi{the} maximum value of the depths of the co-located CTU in the base view and its eight neighboring CTUs are used to limit \hadi{the} maximum depth value of the current CU.}  

{\ekrem{Wang~\etal~\cite{MV-HEVC2} propose an early termination depth splitting algorithm for MV-HEVC.} For a base view, three spatially neighboring CUs (Left (L), Upper (U), and Upper right (UR)) as well as co-located CTU (T) \hadi{are used} to determine \hadi{the} maximum depth value for the current CU. For dependent views, co-located CTU in the base view, called interview CU (I) is used in addition to \hadi{the} above-mentioned neighboring CUs \ekrem{and $D_{pre}$, which will be used in the final decision, is determined as in Eq.~\ref{eq:mvhevc2}.}
\begin{equation}
\label{eq:mvhevc2}
D_{pre} = 
    \begin{cases}
max\{D_L,D_U,D_{UR},D_T\} & \text{Base View} \\
max\{D_L,D_U,D_{UR},D_T,D_I\} & \text{Dependent View} \\
\end{cases}
\end{equation}
 
To avoid wrong decisions, another condition is used to determine \hadi{the} maximum depth value of the current CU (C) which uses motion vector information of three spatially neighboring CUs as \ekrem{shown in Eq.~\ref{eq:mvhevc2_2}.}
\begin{equation}
\label{eq:mvhevc2_2}
\small
D_{max} = 
    \begin{cases}
D_{pre} & \text{$MV_c = MV_L || MV_c = MV_U || MV_c = MV_{UR}$} \\
3 & \text{others} \\
\end{cases}
\end{equation}
}

{\ekrem{Khan~\etal~\cite{MV-HEVC9} use the maximum depth level of interview co-located CTU and its eight spatially neighboring CTUs as a threshold to stop splitting of the current CTU for MV-HEVC.}}%

{\ekrem{Zhang~\etal~\cite{3D-HEVC6} use a two-step strategy for 3D-HEVC to speed up the search for the optimal depth level for CUs for dependent views.} First, an early merge mode decision is made to avoid unnecessary searching of intra and inter modes. If interview CTU and its four immediate neighboring CTUs in the base view are \hadi{en}coded as \hadi{a} merge mode and the RD cost of \hadi{the} skip mode is less than $2N\times 2N$ merge mode for the current CU, only \hadi{the} merge mode for the current CU is searched. Second, \ekrem{CU splitting is terminated if the following two conditions are satisfied: \begin{enumerate*}[label=\textit{\alph*)}]
\item depth of the current CU is equal to or larger than the maximum depth of the interview CTU and its four immediate neighboring CTUs in the base view, and
\item skip mode is selected as the best prediction mode for the current CU after checking all the possible prediction modes
\end{enumerate*}.}}%

{\ekrem{Wang~\etal~\cite{3d-HEVC8} propose a depth range selection algorithm for dependent texture views in 3D-HEVC.} First, split complexity (SC) of left (L) and upper (U) spatially neighboring CTUs, temporally co-located CTU (C) and co-located interview CTU (I) is calculated \ekrem{using Eq.~\ref{eq:3dhevc8}.}

\begin{equation}
\label{eq:3dhevc8}
    SC_i = 
    \begin{cases}
    \frac{1}{256}\sum_{j=0}^{256}d_j & \text{if $depth_{max}=0,1,2$} \\
    \frac{1}{128}\sum_{j=0}^{256}d_j & \text{if $depth_{max}=3$} \\
\end{cases}
\end{equation}
    
Where $d_j$ represents depth level of $4\times 4$ CUs inside a CTU and $depth_{max}$ is the maximum depth of the CTU. Second, $SC$ of the current CTU ($SC_{pre}$) is predicted \ekrem{using  Eq.~\ref{eq:3dhevc8_2}.}

\begin{equation}
\label{eq:3dhevc8_2}
    SC_{pre} = w_C SC_C+w_I SC_I+w_L SC_L+w_U SC_U 
\end{equation}
where $w_i$ is a weighted value. Then a pre-determined threshold is used to find the depth range for the current CU.
}%

{Silva~\etal~\cite{3D-HEVC3} propose a CTU partitioning algorithm for the texture CTUs based on the interview correlation for 3D-HEVC. Independent views are encoded by simulcast HEVC, while dependent views are encoded using the depth information of the corresponding independent view. For the current CU, the corresponding CU in the independent view is determined by a disparity vector~$d(Y)$~\cite{disparityVector} which is calculated as shown in Eq.~\ref{eq:3dhevc3}.

\begin{equation}
\label{eq:3dhevc3}
    d(Y) = \frac{f\times l}{255} \left(\frac{1}{Z_{near}}-\frac{1}{Z_{far}}\right) Y+ \frac{f\times l}{Z_{far}}
\end{equation}
where $l$ is \hadi{the} distance between two adjacent cameras, $f$ is the focal length, $Y$ is the depth view value, and Z values define \hadi{the} depth range of the scene. Positions of two CUs are \ekrem{mapped using Eq.~\ref{eq:3dhevc3_2}.}

\begin{equation}
\label{eq:3dhevc3_2}
    x_{Ind} = x_{Dep} + d(Y)
\end{equation}
where $x_{Ind}$ is position of the current CU in dependent view and $x_{Dep}$ is the position of the same CU in the independent view. Finally, CU depth map is refined.}%

{Mora~\etal~\cite{3D-HEVC4} exploit the correlation between CU depth splitting in texture and depth for 3D-HEVC. For the cases that depth is encoded before the texture, the texture quadtree starts from the coded depth quadtree. Otherwise, the depth quadtree is limited to the maximum and minimum depth values of the texture quadtree.}%

\subsubsection{Multirate}
{\ekrem{Schroeder~\etal~\cite{MR_BlockStruct} first encode the video sequence at the highest bitrate and then use encoded representation as the reference for encoding dependent representations.} As CTUs tend to take larger depths at higher bitrate/quality representations than the lower bitrate/quality representations, the maximum depth value of a CTU in the reference representation is used to limit splitting co-located CTUs in the depended representations.}%

{The same idea has been extended in~\cite{MR_MultiRate} to be used by dependent low-resolution representations.  
As CTUs do not contain the same area in different resolutions, the corresponding area ($A$) for the current CTU in a dependent low-resolution representation is found in the reference representation. \ekrem{Thereafter, the percentage of CTUs ($p_i$) encoded at depth (less than or equal to) $i \in \{0,1,2\}$ is measured.} If $p_i$ (starts from $i=0$) is greater than or equal to a threshold $\theta$, the current CU is not split, and the process moves on to the next CTU.}%

{Ideas in~\cite{MR_BlockStruct} and~\cite{MR_MultiRate} have been combined in~\cite{MR_EffMR} to introduce an algorithm that is used for both dependent low-quality and low-resolution representations.}%

{
\ekrem{Amirpour~\etal~~\cite{dcc2020} propose using both the highest and the lowest quality representation to limit CU depth search range.} First, the highest quality representation is encoded using \ekrem{standard} HEVC, then CTU information obtained in this encoding is used to encode the lowest quality representation similar to \cite{MR_BlockStruct}. Finally, these two representations are used to put both a lower and upper bound for the CU depth search range while encoding the immediate representations. Maximum depth value in the co-located CTU in the highest representation is used as the upper bound, and minimum depth value in the co-located CTU in the lowest representation is used as the lower bound for the CU depth search range. 
}

\revision{In another study, Amirpour~\etal~\cite{MMM21} focus on improving the parallel encoding performance for multi-rate encoding. Different quality representations are chosen as the reference representation to evaluate the performance of parallel encoding. The middle-quality representation is used as the reference representation based on these experiments.}

\subsection{Inherent Approaches}

Inherent approaches use information available within the CTU to give the final decision about CTU partitioning. 

{\ekrem{Kim~\etal~\cite{H_FastRD} decide CU early termination based on the RD cost of a CU.} If it is below a threshold, then current CU is selected as the best one and search process is stopped. A pre-calculated threshold for each CU depth level is considered which is calculated for each CU with size $N\times N$ as $TH_{N\times N} = A_{N\times N} ~ e^{(W_{N\times N}QP)}$.}

\revision{Fu~\etal~\cite{3D-HEVC5} utilize Depth Intra Skip (DIS) for depth map coding, which directly uses reconstructed value of spatially neighboring CUs to represent the current CU for 3D-HEVC.}%

\revision{Fu~\etal~\cite{Corner3D} found that CUs with multi-directional edges have a high correlation with the distribution of the corner points (CPs) in the depth map, and an efficient encoding approach for 3D-HEVC was proposed. Since corner points can indicate multi-directional edges around their neighborhoods, they are used to guide the CU/PU partitioning decisions. The main goal here is to split CU/PU until it contains only one major pattern since no single prediction mode can provide accurate prediction if there are multi-directional edges in a CU/PU. Different approaches are proposed for CU decision, PU decision, and mode decision for 3D-HEVC, which are all based on using CPs.}

\ekrem{In the remainder of this section, we discuss about rest of the inherent approaches by categorizing them into Bayesian and texture complexity based approaches.}

\subsubsection{Bayesian Approaches} 
Bayesian rule based CTU partitioning is an another approach applied in the literature. \ekrem{Bayesian rule is defined in Eq.~\ref{eq:bayes}.}
\begin{equation}
\label{eq:bayes}
p(x\vert y) = \frac{p(y\vert x)p(x)}{p(y)}
\end{equation}
where $p(x\vert y)$ is \textit{posterior}, $p(y\vert x)$ is \textit{likelihood}, $p(x)$ is \textit{prior} and $p(y)$ is \textit{evidence}. In the context of CTU depth decision, CU split decision is seen as the \textit{posterior} and different Bayesian approaches are used to estimate it. 

\ekrem{Shen~\etal~\cite{H_Bayesian} propose selecting certain features, \eg RD cost and inter-mode prediction error, and then use them to minimize the Bayesian risk which is further included in the CU split decision.} CU split decision is defined as a two-class classification problem where split class is $w_{s}$ and non-split is $w_{n}$. \ekrem{Bayesian Risk $R$ of splitting a CU is defined as shown in Eq.~\ref{eq:bayes1}.}

\begin{equation}
\label{eq:bayes1}
    \begin{split}
        R(w_{s}\vert F) = C_{s, n}P(w_{n}\vert F) \\
        R(w_{n}\vert F) = C_{n, s}P(w_{s}\vert F)
    \end{split}
\end{equation}
where $C_{s, n}$ is RD cost of splitting CU when the ground-truth is non-splitting and $C_{n, s}$ is vice versa. $P(w_{s}\vert F)$ is the conditional probability of splitting a CU given the feature vector $F$ which can be re-written using the Bayesian rule \ekrem{in Eq.~\ref{eq:bayes1_2}.}
\begin{equation}
\label{eq:bayes1_2}
    P(w_{s}\vert F) = \frac{P(F\vert w_{s})P(w_{s})}{P(F)}
\end{equation}
where $P(w_{s})$ is the prior probability of splitting a CU. $P(F)$ is constant so it can be ignored and if we replace $P(w_{s}\vert F)$ in \ref{eq:bayes1} with $P(F\vert w_{s})P(w_{s})$, final Bayesian threshold can be written as \ekrem{shown in Eq.~\ref{eq:bayes1_3}.}
\begin{equation}
\label{eq:bayes1_3}
    \frac{P(F\vert w_{s})}{P(F\vert w_{n})} < \frac{C_{s, n}P(w_{n})}{C_{n, s}P(w_{s})}
\end{equation}

$P(F\vert w_{i}), i \epsilon \{n,s\}$ is calculated offline for each QP and resolution settings. Overall, for each CU depth decision, feature vector $F$ is extracted and \hadi{the} offline calculated threshold is used for the decision. CU is non-split when this condition is met, otherwise it is split.

\ekrem{Lee~\etal~\cite{H_BayesianFast} propose a similar approach that benefits from Bayesian rule and models the CU decision problem as a two-class classification.} For both CU depth skip decision and early termination, thresholds are used and calculated using the Bayesian rule. Statistical probabilities for defining thresholds are updated at predefined intervals to keep the approach relatively content-sensitive. 

\ekrem{Xu~\etal~\cite{H_BayesianMBS} use the motion of co-located CTUs for early CU depth search termination along with a Bayesian risk-based discriminant function for detecting skip modes.} CU depths in the co-located CTUs are directly used for early split decision of depth $0$ and depth $1$. Also, average and predicted motion vectors are used to calculate the motion diversity of the co-located CTU, and it is used for early termination. If the co-located CTU is split and motion diversity is high, then the current CTU is early split, and the mode search is skipped. Finally, they model skip mode detection as a two-class classification and use Bayesian risk to approximate it similar to~\cite{H_Bayesian, H_BayesianFast}. Skip mode detection is also used in the early CU depth search termination. It is modeled as a two-class classification problem again, and Bayesian risk-based skip mode detection is used in the decision process. All the statistical parameters in Bayesian approaches are updated periodically to preserve the content sensitivity of the approach.

\ekrem{Jiménez-Moreno~\etal~\cite{H_BayesianAdaptive} propose using Likelihood ratio test (LRT) in Eq.~\ref{eq:bayesianadaptive} for every CU depth level .}
\begin{equation}
\label{eq:bayesianadaptive}
    \setstackgap{L}{.5\baselineskip}
    \frac{P(x\vert depth^*>d)}{P(x\vert depth^*=d)} \Vectorstack{> <} \frac{C_{s, n} - C_{n, n}}{C_{n, s} - C_{s, s}}
\end{equation}
where $x$ is the input feature, $d$ is the current depth, $depth^*$ is the optimal depth, $C_{s, n}$ is cost of splitting CU when the correct decision is not splitting, $C_{n, s}$ is vice versa, $C_{s, s}$ is cost of splitting the RD cost where the correct decision is also splitting and $C_{n, n}$ is vice versa. If the ratio is lower, then CU is not split otherwise it is split. \ekrem{Here RD cost of splitting CU is used as the the feature, $x$}. Statistical properties that are used in the LRT are first extracted by analyzing number of sequences offline \ct{and} they are then updated online to better adapt \ct{to} the changes in the content. 

\ekrem{Lee~and~Jeong~\cite{H_BayesianQDA} use the pixel variance difference between CUs and sub-CUs for measuring local complexity.} Also, a predicted depth value is obtained using weighted information of neighboring CTUs. These two attributes are then combined to be used in early split decision. Finally, a Bayesian decision rule based Quadratic Discriminant Analysis (QDA) is used to classify early termination for CUs. All the thresholds in this method are updated by \ct{an} online learning approach using statistical properties that are extracted during encoding. 

\subsubsection{Texture Complexity}
Texture complexity of the CTUs are also exploited commonly in the literature using different approaches to detect motion. 

{\ekrem{Jamali~and~Coulombe~\cite{gradient} propose an intra coding method based on the global and directional gradients.} Based on the accuracy of the prediction in the current depth, CUs are classified into two categories: \textit{split} and \textit{non-split}. The classification is solved by using global gradient and \hadi{the} mean of gradient amplitudes (MGA) is \ekrem{calculated using Eq.~\ref{eq:gradient}.}
\begin{equation}
\label{eq:gradient}
    MGA = \frac{1}{n} \sum_{i}^{}\sum_{j}^{}|G_X(i,j)| +|G_Y(i,j)|
\end{equation}
where $G_X$ and $G_Y$ are horizontal and vertical Sobel gradient components with $3\times3$ convolution masks applied at each pixel of the current CU. As the CUs with larger $MGA$ have more details, they tend to split while splitting CUs with smaller MGA are stopped at the current depth level. To apply the impact of $QP$ and CU size, $f1$ is \ekrem{calculated using Eq.~\ref{eq:gradient_2} and it is used to determine CU class.}

\begin{equation}
\label{eq:gradient_2}
    f1 = \frac{MGA}{\alpha}-QP
\end{equation}

where $\alpha$ is related to the CU size. For $f1$ values less than a threshold splitting process is terminated. 

\ekrem{Moreover, the directional gradient is used for intra mode decision}. Each CU is classified into a non-split class if the CU intra mode is predicted with high\hadi{er} accuracy at the current level, or a split class, if there is no intra mode, to effectively predict the CU partitioning. By early termination of the splitting process for the non-split class, the encoder saves a considerable amount of time and computations since further splitting would require many RDO computations to find the optimum splitting pattern.}

{\ekrem{Jian~\etal~\cite{H_FastPyramid} indicate that there is a strong correlation between RD cost of a CTU and \hadi{the} corresponding variances of pixel motion vectors.} It is shown that when the motion is strong, blocks tend to take smaller CUs, and when the motion is weak, blocks tend to take larger CUs. Based on these facts, a pyramidal motion divergence (PMD) method is proposed where frames are downsampled to $1/16$, and then their estimated optical flows are used to calculate PMD features. Thereafter, $k$ nearest neighboring-like method is used to predict CU sizes.} %

{This approach is further improved in~\cite{H_PyramidMRF} since calculating PMD was a time-consuming process. Instead, this time variance of absolute difference (VAD) is used as a metric. MVs of neighboring CTUs are used for calculating VAD and Pyramid VAD (PVAD), which is calculated using the downsampled versions. PVAD is used as a feature in the CU split decision problem. Furthermore, the CU split decision is modeled as a Markov Random Field (MRF), and the graph cut method is used to find a decision. The encoded frame is represented as a graph, and CUs are nodes of the graph. Two terminal nodes source $S$ and sink $T$ represent split and unsplit decisions. SVMs are also used in the energy function of the graph to determine the unary term, which is then used for calculating the likelihood of splitting a given CU. Then CU split decision is given based on the minimum cut in the corresponding graph.}%

{A key-point detector that finds high-frequency areas in the image is used to decide on CU depth search range by Kim~\etal~\cite{H_KeypointCU}. The primary motivation behind this approach is that, in HEVC, high-frequency areas are given higher CU depths. Thus, in the proposed method, an adaptive key-point threshold is decided first, and if there are not enough key points in the current CU depth level, further depths are not searched.}

{Nishikori~\etal~\cite{H_VarianceCU} use the variance of the image to determine the characteristic of the region. If the variance is below a predefined threshold, that region is regarded as flat, and the current CU depth is used; otherwise, the CU depth value is increased, and the variance is rechecked.}

\ekrem{Min~and~Cheung~\cite{H_DiagonalVar} propose to use edge complexity to find CU depth sizes.} The edge complexity is calculated in four directions as the luminance difference of the two halves in the corresponding direction. The edge complexity is calculated for all sub-CUs, and if all edge complexity \ekrem{values are smaller than a predefined threshold}, the corresponding CU is not split.

\ekrem{Huang~\etal~\cite{H_CUTexture} use CU texture complexity along with spatially neighboring CTU depth information.} It is calculated by quantizing the variance of the CU into five category levels. CU split decision is given based on the categorization of the CU and information from neighboring CTUs.

\ekrem{Xiong~\etal~\cite{H_SADMECU} propose an approach that designs a new motion estimation (ME) method which can obtain the sum of absolute differences (SAD) costs of both the current CU and sub-CUs.} It also defines an exponential model used to calculate the relationship between motion compensation RD cost and SAD cost. This model is used to calculate a threshold used in CU depth decisions by comparing it with the SAD cost difference.

\ekrem{Song~\etal~\cite{H_DVT} use the discretization total variation (DVT) to calculate CU complexity that is further used in CU depth selection.} If the DVT is high, it means that the current CU is a complex CU that should have higher depth values. 

A Discrete cosine transformation (DCT) based approach is proposed by \ekrem{Liu~\etal~\cite{H_ZeroCU}}. DCT coefficients are checked for early termination. If all DCT coefficients are zero, then the search range is set to 0 and 1 only; otherwise, neighboring depth information is checked. If there is not enough correlation between current CTU and neighboring CTUs, then edge gradient using Sobel edge detector is found and used as the main feature in CU depth decision.

{\ekrem{Ramezanpour~and~Zargari~\cite{ET} define a smoothness parameter for each CU.} If this parameter is lower than a threshold, PU modes are computed for the current CU, and further division is skipped. The smoothness parameter is based on SAD values calculated for horizontal, vertical, right, and left diagonal directions. The variance of these four SADs is defined as the smoothness parameter. }%

{\ekrem{Shang~\etal~\cite{ET2} define edge pixel density $\rho_{edge}$ as in Eq.~\ref{eq:ET2} to represent CU texture complexity.}
\begin{equation}
\label{eq:ET2}
    \rho_{edge} = \sum_{edge}{} /N^2 
\end{equation}
where $\sum_{edge}{}$ is number of edge pixels produced by Canny operator and $N$ is \hadi{the} width of the CU. If $\rho$ is not equal to zero, CU size $64\times 64$ 
is skipped.}%

CU complexity is decided by checking the cost of encoding MVs of the current CU and early search termination is done based on a threshold in \ekrem{\hadi{the} method proposed by Shan~\etal~\cite{H_CUMVBased}.}

\ekrem{Fernández~\etal~\cite{H_TemporalMotion} apply motion estimation on input images and homogeneity analysis is done that is further used to give CU split decision.} \ekrem{Input frames are} analyzed before encoder starts encoding and the process is done on the GPU; thus it does not introduce any overhead for the CPU. After this process, the mean absolute deviation of the motion vectors is obtained for the CU, and the split decision is made based on that. If it is below a certain threshold, then further splits are stopped. 

CTU partitioning decision is made based on the complexity of the CU both in macro and micro levels by \ekrem{Zhang~\etal~\cite{H_Entropy}}. The first frame of the video is analyzed, and the video is categorized into three sizes based on the number of CTU blocks in the frame. If the number is below a threshold, then the number of pixel types is used to determine a rapid decision about CU split. Otherwise, a final decision is given based on the statistical properties, \eg entropy, and texture complexity of CU, and thresholds are set adaptively.

\ekrem{Hou~\etal~\cite{H_texture} calculate the complexity of the CU as $complexity = log(E)$ where E represents the variance of the pixel values in the CU.} Complexity of the current CU is compared with the complexity of the left and top CTUs, and if the complexity of the current CU is smaller than the left and top CTUs, the algorithm stops splitting the CU.%

\ekrem{Cebrián-Márquez~\etal~\cite{H_LookAhead} use a pre-analysis stage called the look-ahead stage.} In the look-ahead stage, the motion information of the sequence is estimated before starting encoding, and this information is later used to guide the CU depth decisions during the encoding process. In this stage, the motion estimation is carried \hadi{out} for each block size and on every reference frame, and the resulting RD costs are stored. Here the same ME algorithm is used as the one in the \ekrem{inter prediction module of standard HEVC} to obtain consistent results. Instead of making a full-motion vector prediction, MVs of spatially neighboring CTUs are used as predictors to speed up the process. The predicted motion information is used for determining a cost function for the given CU, and this cost function is used in determining the final partition. This estimated cost function is calculated using the predicted MV in the look-ahead stage and the distortion rate of the predicted MV compared to the original one. This cost is used as a threshold for a split decision in a bottom-up manner. If the splitting cost is higher than this threshold, then the CU depth is decreased by one until this cost becomes smaller. 

Texture similarity between temporally and spatially neighboring CTUs is used to early terminate CU depth search \ekrem{by Lu~\etal~\cite{H_TextSCC} for HEVC-SCC extension}. The density of the luminance disparity (DLD) is \ekrem{calculated using Eq.~\ref{eq:textscc}, and it is used as the measure.}
\begin{equation}
\label{eq:textscc}
    DLD = \frac{\sum_{x=1}^{W}\sum_{y=1}^{H}\vert I_{cur}(x,y) - I_{col}(x,y)}{W\times H}
\end{equation}
where $W$ and $H$ are width and height of the current CTU, $I_{cur}(x,y)$ and $I_{col}(x,y)$ are luminance intensities of pixel location $(x,y)$ in the current CTU and co-located CTU. If the $DLD$ is smaller than $1$, then the current CTU is categorized as \textit{type 1} otherwise it is categorized as \textit{type 2}. \textit{Type 1} means there is a small variation in luminance compared to the co-located CTU, and the current CTU is expected to have the same CU depth as the co-located CTU. For \textit{Type 1} CU: \textit{(i)} if the depth is smaller than \hadi{that of the} co-located CTU, the CU is further split; \textit{(ii)} if the depth is larger than or equal to \hadi{that of} the co-located CTU and the prediction mode for the co-located CTU is not PLT, then the CU search is early terminated; \textit{(iii)} otherwise, a full-depth search is performed. For \textit{Type 2} CU, spatially neighboring CTUs are also included in the decision process. The maximum and minimum depth among neighboring CTUs are obtained and form a bound for the CU depth search range as $D_{max}$ and $D_{min}$. If the depth of the current CU is smaller than $D_{min}$, the CU is split. If the \ekrem{depth of} current CU is larger than $D_{max}$, the CU search is terminated. Otherwise, a full-depth search is performed.

{\ekrem{Sun~\etal~\cite{H_DirectionVar} use directional variance to measure texture complexity that is further used in CU depth decision.} Directional variance for image $X$ in a given direction $r$ can be written as \ekrem{shown in Eq.~\ref{eq:directionvar}.}
\begin{equation}
\label{eq:directionvar}
    DirVar(X, r) = \frac{1}{N}\sum_{i=1}^{n}\sum_{j=1}^{k_{i}}\left | X_{j} - X_{L(r,i)} \right |
\end{equation}
where $N$ is the number of pixels in $X$, $X_{L(r,i)}$ is the average luminance value along a line with slope $r$ and offset $i$, $X_{j}$ is each pixel in the same line, $n$ is total number of lines, and $k_{i}$ is pixel location in line $i$. 

\ekrem{This approach calculates the sum of variances along a line with a given slope which makes it sensitive to directions, \eg if there are edges along a certain direction, then the directional variance will be larger in that direction.} This allows capturing texture direction information in the image. The set of slopes are determined based on the mode directionality of HEVC and four slopes are selected, \ie $0^{\circ}, 45^{\circ}, 90^{\circ}$, and $135^{\circ}$. Based on the directional variances in these slopes and pre-determined thresholds, the CU is categorized into one of the three groups, \ie homogeneous, complex, and undetermined. Homogeneous CUs are not split, complex CUs are split, and full RDO search is applied for undetermined CUs. Different threshold values are used for different QP values. 
}

{\ekrem{Chiang~\etal~\cite{3D-HEVC1} calculate variance of pixels inside the CU ($Var_{cu}$) and compare it with a pre-defined threshold to determine CU decision for 3D-HEVC. CU is split if $Var_{cu}$ is larger than a threshold, otherwise depth of the CU is compared with the depth of co-located texture CU and depth search is terminated if it is larger or equal.}

\ekrem{Li~\etal~\cite{3D-HEVC2} propose another method for 3D-HEVC and use pre-determined thresholds as well as RD-cost of CU at depth $0$ ($J0$) to determine the maximum depth value ($d_{max}$) for a depth map CTU.} The maximum depth value is determined as \ekrem{shown in Eq.~\ref{eq:3dhevc2}.}
\begin{equation}
\label{eq:3dhevc2}
d_{max} = 
    \begin{cases}
0 & \text{if $J_0 \leqslant Th0$} \\
1 & \text{if $Th0 \leqslant J_0 < Th1$} \\
2 & \text{if $Th1 \leqslant J_0 < Th2$} \\
3 & \text{if $J_0 > Th2$} 
\end{cases}
\end{equation}
}%

\revision{Zhang~\etal~\cite{Prob3D} use two techniques to speed up the depth intra-mode decision process in 3D-HEVC: 
\begin{itemize}
    \item A criterion based on the squared Euclidean distance of variances (SEDV) is used to evaluate RD costs of the Depth Modeling Mode (DMM) candidates, based on the the statistical characteristics of variance distributions. 
    \item A probability-based early depth intra-mode decision is proposed to choose only the most probable mode.
\end{itemize}
}

Overall, statistics based methods and common features used in them are summarized in Table \ref{table:heuristic}. 

\subsection{Summary}

Neighboring CTU information is commonly used in the statistics-based approaches since the correlation is vital for such CTUs. The most common approach here is to define a threshold and give the decision based on depth values of neighboring CTUs~\cite{H_NeighbourCU_2, H_ContentFastCU, H_NeighbourCU, H_SeqSim, H_Combined}. Some approaches directly use minimum and maximum depth values found in the neighboring CTUs to limit the search range~\cite{H_CULimitFrame, H_InterCUEn, LF1, LF2, H_FrameAndPU, H_HEVCEff, H_CUDist, H_RoughAccurate}. Additionally, categorizing frames using this CTU information and giving the decision based on the classification is another common approach~\cite{H_complexityControl, H_FrameEff_2}. 
 
\ekrem{HEVC extensions and multirate approaches provide additional neighboring CTUs that can be used in the process.} For MV-HEVC and 3D-HEVC, there is one extra co-located CTU in the interview \ekrem{frame}, and for 3D-HEVC, there is also one extra CTU in the depth frame~\cite{MV-HEVC}. Numerous approaches are proposed that specifically aim to exploit those extra neighboring CTUs in these HEVC extensions~\cite{MV-HEVC3, MV-HEVC1, MV-HEVC2, MV-HEVC9, 3D-HEVC6, 3d-HEVC8, 3D-HEVC3, 3D-HEVC4,  3D-HEVC5, MR_MultiRate, MR_BlockStruct, MR_EffMR, dcc2020}. These methods also follow similar approaches as \ekrem{methods proposed for standard HEVC}. 

Inherent approaches, on the other hand, exploit the information available within the current CTU. 

The bayesian rule is exploited in the context of CTU partitioning. Some approaches focus on minimizing the Bayesian risk of splitting CU by using different feature sets~\cite{H_Bayesian, H_BayesianFast, H_BayesianMBS}. Here, RD cost is an essential feature since it can be correlated with the Bayesian risk. Also, Bayesian decision rules are used to determine CU split decision again RD cost being the main feature here as well~\cite{H_BayesianAdaptive, H_BayesianQDA}.

Calculating texture complexity and using it to determine CTU partitioning is also used commonly in \hadi{the} inherent approaches. This is useful since more complex CTUs tend to have larger depths to achieve better motion compensation. Numerous information sources are exploited here from directional gradients~\cite{gradient}, pyramidal motion divergence~\cite{H_FastPyramid, H_PyramidMRF}, RD cost of encoding CU~\cite{H_FastRD}, \etc~However, the most common approach is to determine the texture complexity of the CTU using the variance of pixels since variance is strongly correlated with the texture complexity of the CTU~\cite{H_VarianceCU, H_CUTexture, H_texture, H_DirectionVar, 3D-HEVC1, 3D-HEVC2}. Moreover, motion vectors are key factors in determining texture complexity as well, and they are also exploited commonly in these methods~\cite{H_PyramidMRF, H_CUMVBased, H_TemporalMotion, H_LookAhead}.

\begin{table*}[pos=t]
\centering
\caption{Statistics Approaches.}
\label{table:heuristic}
\resizebox{\textwidth}{!}{%
\begin{tabular}{l|l|l|l}
\multicolumn{1}{c|}{\textbf{Method}}          & \multicolumn{1}{c|}{\textbf{Approach}}           & \multicolumn{1}{c|}{\textbf{Features}}                       & \multicolumn{1}{c}{\textbf{Mode}} \\ \hline
\cite{H_NeighbourCU_2}                        & Neighbour CTU                                    & Depth Values                                                 & Intra                             \\
\cite{H_ContentFastCU}                        & Neighbour CTU                                    & RD Cost, Prediction mode                                     & Intra                             \\
\cite{H_NeighbourCU}                          & Neighbour CTU                   & Depth Values, Frame Type                                     & Inter/Intra                       \\
\cite{H_SeqSim}                               & Neighbour CTU                                    & Depth Values, RD Cost                                        & Inter/Intra                       \\
\cite{H_Combined}                             & Neighbour CTU                   & Depth Values, RD Cost                                        & Intra                             \\
\cite{H_CULimitFrame}                         & Neighbour CTU                                   & Depth Values                                                 & Inter/Intra                       \\
\cite{H_InterCUEn}                            & Neighbour CTU                   & Depth Values                                                 & Inter                             \\
\cite{LF1,LF2}                                & Neighbour CTU                    & Depth Values                                                 & Inter/Intra                       \\
\cite{H_FrameAndPU}                           & Neighbour CTU                                   & Depth Values, PU Mode                                        & Inter/Intra                       \\
\cite{H_Spatiotemporal}                       & Neighbour CTU                    & Depth Values                                                 & Inter/Intra                       \\
\cite{H_CBFrameSkip}                          & Neighbour CTU                                  & Depth Values                                                 & Inter                             \\
\cite{H_HEVCEff}                              & Neighbour CTU                                   & Depth Values, Motion Vectors, RD Cost, skip Mode             & Inter                             \\
\cite{H_complexityControl}                    & Neighbour CTU                                   & Depth Values                                                 & Inter/Intra                       \\
\cite{H_FrameEff_2}                           & Neighbour CTU                    & Depth Values, Motion Vectors                                 & Intra                             \\
\cite{H_DepthHistory}                         & Neighbour CTU, Frame History                    & Depth Values                                                 & Inter                             \\
\cite{SHVC1}                         & Neighbour CTU, SHVC                    & Depth Values                                                 & Inter                             \\
\cite{SHVC2}                         & Neighbour CTU, SHVC                    & Depth Values, RD Cost                                             & Inter                             \\
\cite{SHVC3}                         & Neighbour CTU, SHVC                    & Depth Values                                                 & Inter                             \\
\cite{SHVC4}                         & Neighbour CTU, SHVC                    & Depth Values, RD Cost                                                 & Inter                             \\
\cite{SHVC5}                         & Neighbour CTU, SHVC                    & Depth Values                                                 & Intra                             \\
\cite{SHVC6}                         & Neighbour CTU, SHVC                    & Depth Values                                                 & Inter                             \\
\cite{SHVC8}                         & Neighbour CTU, SHVC                    & Depth Values, Texture Information                                                  & Inter                             \\
\cite{SHVC9}                         & Neighbour CTU, SHVC                    & Depth Values                                                 & Intra                             \\
\cite{MV-HEVC1}                               & Neighbour CTU, MV-HEVC                    & Depth Values                                                 & Inter                             \\
\cite{MV-HEVC2}                               & Neighbour CTU, MV-HEVC                    & Depth Values, Motion Vectors                                 & Inter                             \\
\cite{3D-HEVC3}                               & Neighbour CTU, 3D-HEVC                    & Depth Values, Camera Properties                              & Inter                             \\
\cite{3D-HEVC4}                     & Neighbour CTU, 3D-HEVC                                    & Depth Values                                                 & Inter                             \\
\cite{MR_BlockStruct, MR_MultiRate, MR_EffMR} & Multirate                                       & Depth Values, Motion Vectors, Encoding Parameters            & Inter                             \\
\cite{dcc2020, MMM21}                         & Multirate                                       & Depth Values, Reference Frames                               & Inter                             \\
\cite{H_Bayesian}                             & Bayesian, Sub-CU                                 & Variance                                                     & Inter/Intra                       \\
\cite{H_BayesianFast}                         & Bayesian, Frame History                          & RD Cost                                                      & Inter                             \\
\cite{H_BayesianMBS}                          & Bayesian, Neighbour CTU                          & Motion Vectors, Depth Values, RD Cost                        & Inter                             \\
\cite{H_BayesianAdaptive}                     & Bayesian                                         & Depth Values, RD Cost                                        & Inter/Intra                       \\
\cite{H_BayesianQDA}                          & Bayesian, Neighbour CTU, Sub-CU                  & Variance, Depth Values, RD Cost                              & Intra                             \\
\cite{H_FastPyramid}                          & Texture Complexity                           & Motion Vectors                                               & Inter                             \\
\cite{H_PyramidMRF}                           & Texture Complexity, Sub-CU                           & Motion Vectors                                               & Inter                             \\
\cite{H_KeypointCU}                           & Texture Complexity                                & Edge Complexity                                           & Intra                             \\
\cite{H_VarianceCU}                           & Texture Complexity                                & Variance                                                     & Intra                             \\
\cite{H_DiagonalVar}                          & Texture Complexity, Sub-CU                        & Variance, Edge Complexity                                    & Intra                             \\
\cite{H_CUTexture}                            & Texture Complexity, Neighbour CTU & Depth Values, DCT Coefficients                               & Intra                             \\
\cite{H_SADMECU}                              & Texture Complexity, Sub-CU                        & RD Cost, SAD                                           & Inter                             \\
\cite{H_DVT}                                  & Texture Complexity, Neighbour CTU                 & Depth Values, RD Cost                                        & Intra                             \\
\cite{H_ZeroCU}                               & Texture Complexity, Neighbour CTU                 & Depth Values, Variance, Texture Complexity, DCT Coefficients & Inter                             \\
\cite{ET}                               & Texture Complexity                 & Variance, SAD, QP & Intra                            \\
\cite{ET2}                               & Texture Complexity, Neighbour CTU                 & Edge Complexity, Depth Values & Intra \\
\cite{H_CUMVBased}                            & Texture Complexity                                & Motion Vectors, RD Cost                                                      & Inter                             \\
\cite{H_TemporalMotion}                       & Texture Complexity, Frame History                 & Motion Vectors                                               & Inter                             \\
\cite{H_Entropy}                              & Texture Complexity                                & Variance                                                     & Intra                             \\
\cite{H_texture}                              & Texture Complexity, Neighbour CTU                 & Variance                                                     & Intra                             \\
\cite{H_LookAhead}                            & Texture Complexity,  Frame History, Neighbour CTU & Depth Values, Motion Vectors                                 & Inter                             \\
\cite{H_TextSCC}                              & Texture Complexity, Neighbour CTU, HEVC SCC & Depth Values, Luminance Values, Encoding Mode                & Intra                            \\
\cite{3D-HEVC1}                               & Texture Complexity, 3D-HEVC                                & Variance, Depth Values                                       & Intra                             \\
\cite{3D-HEVC2}                               & Texture Complexity, 3D-HEVC                                & Depth Values, RD Cost                                         & Intra \\
\cite{Prob3D}                               & Texture Complexity                                & Depth Values, RD Cost                                         & Intra       
\end{tabular}
}
\end{table*}

\section{Machine Learning Based Approaches}
\label{sec:Machine}
Following the recent success of machine learning methods in numerous fields, ML based approaches have also been proposed for CTU partitioning. In this section, \hadi{the} approaches are categorized into two main groups, \hadi{namely} \textit{(i)} traditional and \textit{(ii)} deep learning methods. Traditional methods such as support vector machine and random forests use manually selected features for training\hadi{,} while deep learning methods extract useful features during the \ct{training phase from the data}.

\subsection{Traditional Methods}
In traditional machine learning methods, the first and most important step is feature extraction. These methods \ct{typically} use handcrafted features. One advantage of traditional ML methods over deep learning counterparts is their low complexity. \ct{Additionally}, it is easier to interpret the results of traditional methods. Another advantage over deep learning methods can be seen when the data size is small since these methods are not dependent on the vast amount of data contrary to deep learning methods. In this section, we categorize traditional methods into three groups based on the main method used in the \ct{respective} approach: \textit{(i)} support vector machines (SVM), \textit{(ii)} Random forests (RF), and \textit{(iii)} bayesian learning.

\subsubsection{Support Vector Machines}
\ct{The m}ain idea behind Support Vector Machine (SVM)~\cite{SVM} is to construct a high dimensional hyperplane that classifies data points and maps the given input to the hyperplane for classification. The main goal here is to output a hyperplane that gives minimum classification error. SVM is a supervised learning method meaning that correct labels are available for the entire training dataset. SVMs are usually used for classification tasks, but they can also be used for regression. SVMs are commonly used for CTU partitioning. 

Specifically, \ekrem{Xue~\etal~\cite{ML_SVM} model CU splitting as a binary classification problem and use SVM to solve it.} Three different SVMs are used, one for each depth level, and they are trained offline using the mean squared error (MSE) and \hadi{the} number of encoded bits (NEB), which are extracted from the CUs with different depth levels.

\ekrem{Shen~and~Lu~\cite{ML_WSVM} assign weights to \hadi{the} training samples based on the RD loss increase caused by the misclassification of the sample to reduce the effect of outliers and misclassification.} 
Based on this approach, the CU depth prediction process is performed before checking depth levels to reduce time-complexity \ekrem{by Zhang~\etal~\cite{ML_FlexSVM}.} An initial\hadi{ly} weighted SVM is used to predict CU depth level distribution for the CTU, and the search process is shaped based on the result.

\ekrem{Liu~\etal~\cite{ML_FlexSVMImp} utilize separate weighted SVMs for each depth level and use features related to texture complexity for training SVMs.} Texture complexity of the CUs is measured using Sobel filter on luminance values of pixels and used as \hadi{a} feature along with QP. CU splitting is modeled as three-class (\textit{complex CU, homogeneous CU, uncertain CU}) classification problem. 
If the CU is classified as \hadi{a} complex CU, it is further split. If it is classified as \hadi{a} homogeneous CU, splitting is stopped, and for uncertain CUs normal encoding process is applied.  

Two linear SVMs are used for CU split and early CU termination decisions \ekrem{by Zhang~\etal~\cite{ML_LSVM}.} Depth difference and Hadamard transform-based (HAD) cost ratio between current and neighboring CTUs are used for early CU split decision. RD cost is also used along with these two features for early CU termination.

\ekrem{Liu~\etal~\cite{ML_ANNSVM} use an artificial neural network (ANN) to determine weights for outputs of different SVMs.} 
SVMs trained with variance information are given the highest weights, and SVMs trained with neighboring CTUs depths, and pixel differences are given \hadi{the} equal weights based on the ANN results.

\ekrem{Zhu~\etal~\cite{ML_FuzzySVM} use fuzzy SVM to directly predict CU depths and model\hadi{s} CTU partitioning as \hadi{a} multi-class classification.} 
Three SVM-based classifiers are used that are trained with manually selected features and updated periodically. 
The SVM can also predict uncertainty for those examples that are not confident, meaning they are in the risk area, and in these cases, the standard HEVC search process is used. 
\ct{The r}isk area is determined by RD cost optimization. During \ct{the} training phase, each sample is weighted by fuzzy SVM so that the outliers do not reduce the overall accuracy. After \ct{the} classifiers are trained, \ct{the} CU structure of \ct{the} remaining frames in the GOP is directly predicted. If the sample is in the risk area, then \ct{standard} HEVC search is applied. 

Two SVMs are used for CU decisions by \ekrem{Zhang~\etal~\cite{ML_DoubleSVM}.} \ct{The} SVM is trained offline and outputs three decisions, \ie split, non-split, or full search. If the output of the first classifier is full search, \ct{the} second classifier is used, which is trained online using the data from previous\hadi{ly} encoded frames to further predict \hadi{the} CU size. 

{\ekrem{Erabadda~\etal~\cite{ML_TwoLayerSVM} use weighted SVMs that are trained online during encoding cycle in \hadi{a} two layered structure to determine CU depth early termination.} 
In \ct{the} the first level, two SVMs are used for each CU depth, and training data is collected using default RD optimization of HEVC. Features are selected as texture complexity, RD cost of the current depth, and context information. \hadi{The} weights for these SVMs are calculated using precision scores of split decisions. CUs that are not categorized by the first layer SVMs are passed to the second layer. In the second layer, there is only one SVM per CU depth level, and training data collection is half of the amount compared to the first layer. Weight calculation this time is done by F-score of CU split decisions. In both levels, SVMs are re-constructed after some period to induce some context adaptivity to the SVMs. \ct{Additionally}, they use complexity control parameters to control \hadi{the} number of CUs that have reached the second level to be decided by exhaustive RD optimization.}

{A two stage algorithm for CTU partitioning is proposed \ekrem{by Erabadda~\etal~\cite{ML_TwoLayerSVMInter}.} 
In \ct{the} first stage, offline trained SVMs are used to make the CU splitting decision, and the second stage is used to apply the decision of the SVMs. 
If the decision is split, then the current level CU calculations are skipped. SVMs are not used for depth $2$ since their usage decreased performance significantly during the experiments. Depth information from neighboring CTUs and co-located CTU, QP value, RD-Cost, and texture information are used as features for SVMs. A different set of features is used for depth $0$ and depth $1$ decisions based on their F-scores.}

\ekrem{Xue~\etal~\cite{ML_SVMSCC} model the CTU depth decision for HEVC-SCC as \hadi{a} binary classification task and train separate classifiers for each depth level.} \ct{In particular}, L1-loss linear SVM is used as \hadi{a} classifier. Due to imbalance in the training dataset for SCC, \eg split decisions are much larger than non-split, \hadi{an} ensemble learning approach is adopted. Random subsets are generated from the training set\hadi{,} which are then used for training classifiers, and outputs of classifiers are given weights for the voting function. \ct{The m}ean absolute deviation of luminance values for CU is used to measure texture complexity. However, since text regions are common in screen content, \ct{the} range of luminance values in each CU is quite narrow. To overcome this problem, \ct{the} number of luminance components and \ct{the} range of luminance values in the current block are also used as additional features. Bitrate, RD cost, and QP of planar mode are also used along with \ct{the} depth values of left and up CTU. For CU depth decision, \ct{the} first planar mode is checked, and a feature vector is obtained. After that, a classifier is called, and the decision is made according to the output. If the output is a certain split or a certain non-split, these decisions are applied; otherwise, the full search is conducted.

\subsubsection{Random Forests}
Random forests (RF)~\cite{RandomForest} are also commonly used in the literature. Random forests consist of many decision trees. In decision trees, the prediction is made by gradually limiting the search range of the problem. At the root, a general elimination is done based on a general condition, and once it is moved deeper in the tree, the conditions become more and more specific, thus limiting the search range. A decision tree can have multiple output branches, and the goal is to find the output that minimizes the error. In RF, multiple decision trees that are trained with random training samples and use random feature sets are combined to produce the final output, which is obtained by combining \ct{the} weighted output of \ct{the} decision trees.

\ekrem{Duanmu~\etal~\cite{ML_DecisionTree} propose a method for HEVC-SCC which uses decision trees for CU decision.} \ct{In the beginning}, statistical properties are processed to extract useful features for the decision tree, then a decision tree is trained to classify block type for screen content. Following that, another decision tree is used to decide whether or not to split a CU further. Features that capture texture complexity are used to train decision trees.

\ekrem{Ruiz-Coll~\etal~\cite{ML_DataTree} use a decision tree to determine early termination for CU depth search.} Low complexity features are used to train trees, and they are only used for CU depths $0$ and $1$. 

CU splitting for HEVC-SCC is modeled as a two-class classification problem by Yang~\etal~\cite{ML_DecisionSCC}, and decision trees are used to solve it. The following features are used: variance of luminance of the current CU, maximum gradient magnitude of the current CU obtained by using Sobel operator, CU depth level, information mode of spatially neighboring CUs, and RD cost of CU. A separate classifier is trained for each depth level, and each classifier is trained offline using frames from several different sequences. 

\revision{Fu~\etal~\cite{DecTree3D} propose a decision tree-based approach for 3D-HEVC. Texture and variance information within the CUs and the decision from neighboring CUs are used as features to train decision tree classifiers. A separate set of features and classifiers are used for different encoding decisions. Decision trees are used for outputting skip or check decisions for Intra 2N$\times$2N mode, Intra N$\times$N mode, and CU splitting.}

\revision{Chan~\etal~\cite{Overview3D} provide an overview of depth map coding approaches for 3D-HEVC along with proposing a decision-tree-based depth coding algorithm. Available fast PU and CU size decision algorithms for 3D-HEVC in the literature are presented first, and a learning-based algorithm is proposed. Depth intra skip (DIS) mode is performed first, and the view synthesis optimization (VSO) cost of this process is used as one of the features in the decision-tree classifier. The current CU is classified into one of the three classes (\ie \textit{Skip CU}, \textit{Terminate CU}, and \textit{Uncertain CU}). Based on the classifier output, the current CU is split, CU search is terminated, or the original depth intra coding flow is applied.}

\ekrem{Du~\etal~\cite{ML_RForest} use random Forest to predict CU depth level.} CU depth range is determined using CU depths of neighboring blocks, and each block is given different weights based on their location, \ie left and top blocks have higher weight\hadi{s} than top-left and top-right blocks. If the depth range is $0$ or $1$, a random forest classifier is used to give \ct{the} final decision. For the remaining depth levels (\ie $2$ and $3$), a standard HEVC RDO process is applied.

\ekrem{Tahir~\etal~\cite{ML_RForestFeatureExtr} use three different random forest classifiers for skip mode, CU split, and TU split decisions.} Several features are rank-ordered using a filtering-based approach to find the optimal number of features for each of the three classifiers. This results in 10 features for skip mode, 9 features for CU split, and four features for TU split decisions. Skip mode classification followed by CU split decision and TU split decision is made for each CU level. 
The multirate algorithm proposed by De Praeter~\etal~\cite{MR_MLBased} uses information of blocks in the fully encoded representation that is in the same location as \ct{the} current block to predict the block structure of the current block. Then a random forest is used to predict which information will be used for encoding \hadi{the} remaining representations. Features are chosen as the variance of the transform coefficients, motion vector variance, and information about \ct{the} block structure. 

\ekrem{Bubolz~\etal~\cite{MR_RFHBR} extract features for training the random forest classifier \hadi{of} high bitrate video then train a random forest classifier for low bitrate representations.} \ct{The c}lassifier gives the decision of whether to stick with the same depth level as high bitrate representation or further split the CU for current representation for each depth level. This method mainly focuses on transrating processes.

\subsubsection{Bayesian Learning}
The bayesian rule is also used within machine learning based methods. Different from the Bayesian methods in the statistics based approaches, these methods decide parameters for Bayesian rule with online or offline training. 

\ekrem{Kim~and~Park~\cite{ML_BayesianOnline} choose training pictures based on scene change to update statistical parameters which are used for a Bayesian classifier that decides whether to search further depths for the given CU.} An offline-learning-based loss matrix calculation is also used to improve the decision process further. After obtaining \hadi{the} thresholds using these two approaches, early CU termination is done by calculating posterior cost and comparing it with thresholds. 

Another online learning based Bayesian Decision based Block Partitioning (BDBP) method is proposed by \ekrem{Yao~\etal~\cite{ML_BayesianOnlineGMM}.} \hadi{The} frames in the video are divided into two groups, \ie online learning and fast prediction, based on scene change. 
A Bayesian risk threshold is defined using a Gaussian Mixture Model (GMM), and it is used for the split decision. \hadi{The} scene change is detected using the gray level difference between \hadi{the} frames. At each scene change, an online learning period of six frames is started because the threshold is no longer valid.  \hadi{During} the online learning period, \hadi{the} CU decision is done by \ekrem{standard} HEVC and statistical data is collected. When the online learning period is over, the fast prediction period is restarted, and the threshold is decided in the first frame of \ct{the} fast prediction period using Expectation-Maximization (EM) algorithm, and EM is initialized with the K-Means method to avoid falling into a local minimum.

\ekrem{Chen~and~Lu~\cite{ML_BayesianProgressive} use an online progressive three-class Bayesian classifier along with a general Bayesian classifier.} \ct{The f}irst classification is done by a three-class classifier using features from \ct{the} current CU and \ct{the} neighboring CTUs. If the first classifier gives an indistinct decision for the CU, a second general Bayesian classifier is used to give \ct{the} final decision. Both classifiers are trained online, and parameters are updated periodically. 

CTU partitioning is modeled as binary classification problem by \ekrem{Kuang~\etal~\cite{ML_BayesianSCC}} for HEVC-SCC. RD cost of the optimal mode for \ct{the} current CU depth level, $J$, is used as a feature which is defined as \ekrem{$J = D_{SSE} + \lambda \times R$} where $D_{SSE}$ is sum of squared errors between \ct{the} current CU and the predicted CU, $R$ is \ct{the} bit cost, and $\lambda$ is \ct{the} Lagrange multiplier \ekrem{which is defined in Eq.~\ref{eq:bayesianscc}.}
\begin{equation}
\label{eq:bayesianscc}
    \lambda = C \times \left(\frac{QP - 12}{3.0}\right)^2
\end{equation}
where $QP$ is \ct{quantization parameter} value and $C$ is a parameter determined by the picture type and coding structure. They calculate $J$ directly from \ct{the} SCC encoder\ct{, thus} no overhead is introduced. For the CU depth decision, CUs are categorized into three groups based on their optimal modes, \ie Cintra, IBC, and PLT. Bayesian rule \ekrem{in Eq.~\ref{eq:bayesianscc_2}} is used for determining early CU termination:
\begin{equation}
\label{eq:bayesianscc_2}
    P_{group_{n}, d}(w_{j}\vert J) = \frac{P_{group_{n}, d}(J\vert w_{j})P_{group_{n}, d}(w_{j})}{P_{group_{n}, d}(J)}
\end{equation}
where $j\epsilon \{split, notsplit\}$, $group$ is determined by \ct{the} optimal encoding mode, $P_{group_{n}, d}(w_{j})$ is the prior of $w_{j}$, and $P_{group_{n}, d}(J)$ is the probability density of $J$. Both of these probabilities are obtained using statistical information during the online-learning phase. Probability density of $J$ is obtained \ekrem{using the formula in Eq.~\ref{eq:bayesianscc_3}.}
\begin{equation}
\label{eq:bayesianscc_3}
    P_{group_{n}, d}(w_{j}\vert J) = \sum_{w_{j}} P_{group_{n}, d}(J\vert w_{j})P_{group_{n}, d}(w_{j})
\end{equation}
CUs with smaller $J$ are more likely to belong to $w_{nonsplit}$ class and to be early terminated but there are still exceptions observed during experiments. Thus, this early termination method is only applied for Cintra and IBC modes. Training frames for online-learning phase are chosen based on the scene-change detection which is done by dividing the frame into blocks and using distinct color number in each block. Those frames are then used for updating statistical parameters.  

\subsection{Deep Learning Methods}
Following the recent success of deep learning methods in various tasks, several deep learning based methods are also proposed \ct{for CTU depth decisions algorithms}. Deep learning methods can benefit from \ct{an} excess amount of data thanks to their ability to learn complex non-linear functions of given inputs~\cite{DLLinearComplexity}. Moreover, in deep learning methods, features are extracted implicitly by the model during training which eliminates the need for feature engineering. 
Deep learning methods are more complex compared to traditional methods, hence their training time is longer, and they require more data to work. 
However, if the requirements are met, deep learning methods generally provide better performance compared to traditional machine learning methods for a variety of tasks. 

\subsubsection{Convolutional Neural Networks}
\begin{figure}[pos=t]
\captionsetup{justification=centering,
              singlelinecheck=false,
              font=sf, labelfont=bf} 
\centering
\includegraphics[width=0.48\textwidth]{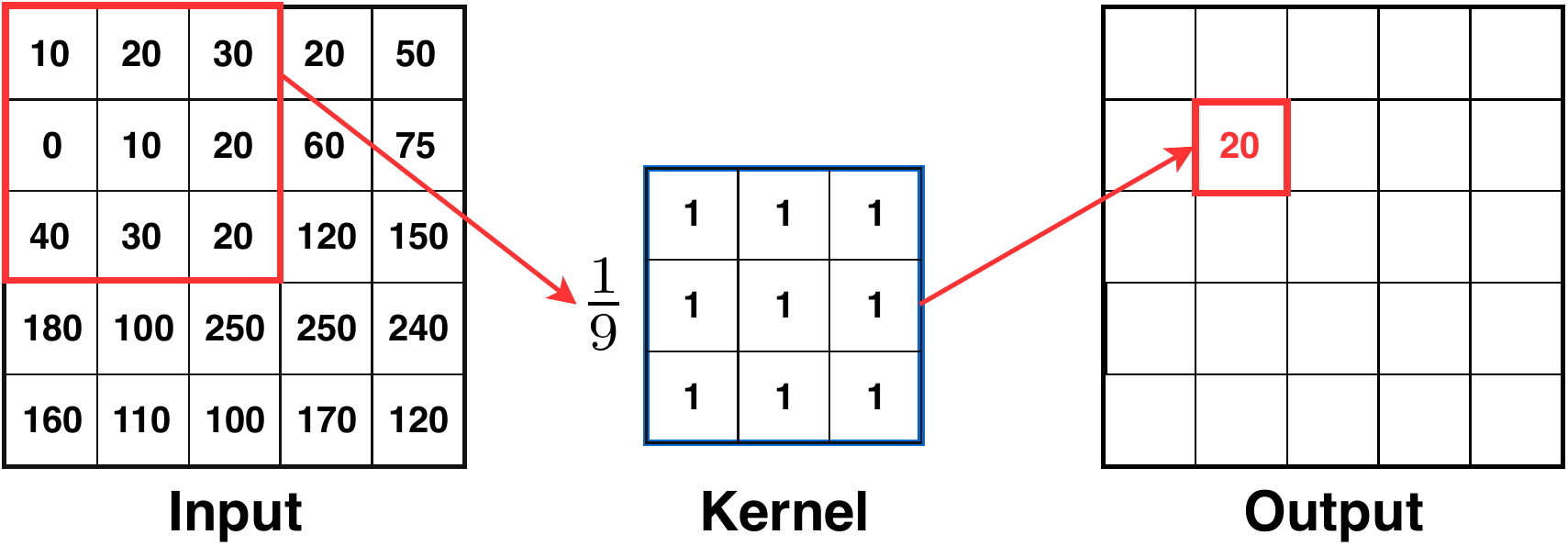}
\caption{Sample Convolution Operation.}
\label{fig:conv}
\end{figure}

Convolutional neural networks (CNNs) are commonly used for the CTU partitioning algorithms since they are good with 2D data such as images. CNNs are a special type of neural networks (NNs) in which mainly convolution operation\hadi{s} is used. Convolution can be written as \ekrem{shown in Eq.~\ref{eq:cnn}.}

\begin{equation}
\label{eq:cnn}
Y[i, j] = \sum_{k, l}f[k, l]X[i-k, j-l]
\end{equation}

where $Y$ is \textit{output}, $f$ is \textit{kernel}, and $X$ is \textit{input}. An example \hadi{of} convolution operation can be seen in Fig.~\ref{fig:conv}. In CNNs, convolution operations are applied consecutively to the given input, thus extracting different features in different layers which makes them extremely useful for capturing important features in the given image.

\ekrem{Kim~and~Ro~\cite{ML_FastCuNN} use a CNN for \hadi{the} CTU partitioning.} CTU partition is searched as a top-down approach from depth $0$ to $2$, and in each level, the CNN decides on whether to split into more depths or \ct{to} stop. \ct{The N}etwork is trained using luminance values in the frame, and manually chosen features are also used as an additional vector to improve \ct{the} prediction accuracy. 

\ekrem{Xu~\etal~\cite{ML_LSTMCNN} propose to use a unique representation for \hadi{the} CU prediction, \ie a hierarchical CU partition map (HCPM) to efficiently represent \ct{the} CU partition of \ct{the} CTU.} Using HCPM and CNN, it can predict \hadi{the} depth decision for the entire CTU in a single inference for intra frame, rather than running prediction for each depth level as in~\cite{ML_FastCuNN}. It also uses long short term memory (LSTM) to reduce the inter-frame CU complexity.

\ekrem{Li~\etal~\cite{ML_DeepCNN} use a deep CNN at each depth level to predict \hadi{the} CU split decision.} \ct{The t}exture of \ct{the} CU is used as the only feature, and CNNs are trained offline using a raw image database. A new dataset for Intra-mode HEVC is constructed to train the network. $2000$ images are encoded with the HEVC reference software, and binary labels are obtained for all CUs. \ct{The d}ataset is then grouped by the QP value and resolution. A separate CNN is trained for each depth level which gives a binary decision for each CU whether to split or not to split.

An asymmetric CNN structure that takes the luminance value of \ct{the} CU as an input is used to predict both CU and PU partitioning decisions \ekrem{by Shi~\etal~\cite{ML_AsymCNN}}. The network is divided into three branches in which two of them have asymmetric kernels in convolutional layers to capture horizontal and vertical features. \ct{Subsequently,} outputs of sub-networks are concatenated in depth level, which is followed by two convolutional layers that extract weight information of features. Different networks are used at each depth level, and PU prediction is modeled as a special case of CU prediction. For PU prediction, softmax activated output of the network is used as confidence level, and PU prediction is made if it is above certain thresholds, which are determined differently for each QP level. 

\ekrem{Zhang~\etal~\cite{ML_TextureCNNMix} use both statistics and machine learning based approaches.} \ekrem{A texture complexity threshold is defined based on QP and depth values.} If the CU is determined to be homogeneous by the threshold, further searches are stopped; otherwise, \ct{the} CU is sent to a CNN for classification. For CNN, kernel sizes are designed dependent on the CU depth, and also neighboring CTU information is included in the fully connected layers. Furthermore, \ct{the} QP is included in the loss function. These changes provide an adaptive structure for CNN. There are different CNNs for each depth level. 

\ekrem{Chen~\etal~\cite{ML_TwoCNNCUPU} use two CNNs for \hadi{the} CU and PU predictions.} CNNs are trained using frames and QP levels as inputs, and the final decision is given based on the predefined threshold. \ct{The} CNN is only used for \hadi{the} split decision when the depth is $3$, which decides if the CU needs to be split further for the PU mode. \ct{The} CNN also considers the edge strength of the CU and QP level when giving the decision. This is achieved by training the CNN with QP level along with input frames. For other depth levels, the \ekrem{standard} HEVC encoding process is applied. 

{A deep CNN, which is designed to classify texture information and propose object locations in the frame, is used to determine CTU partitioning \ekrem{by Kuanar~\etal~\cite{ML_CNNROI_2}.} It first produces \ct{the} ROI for possible objects, then object shape detection is done, and finally, the CTU is classified into one of the four classes (\textit{depth $0, 1, 2, 3$}) based on the combination of object and texture features. For the ROI proposal, predetermined anchor boxes are used, and labeling is done based on a predefined threshold for intersection over union (IoU) scores. For texture features, Fisher vectors are used to aggregate local features that are obtained by the convolutions, and \ct{the} Principle Component Analysis (PCA) is applied to reduce the dimensionality. 
}

{\ekrem{Bouaafia~\etal~\cite{ML_SVMCNN} use both SVM and CNN for the CTU partitioning.} \ct{The} CTU partitioning decision is modeled as \hadi{a} three-level binary classification, each level represents a depth level. Several frames are encoded using the original encoder, and these are used for \ct{the} online training \ct{of} the SVM. \ct{The} SVM is then used for consecutive frames for CU split prediction. This training period is refreshed periodically to make the SVM content-adaptive. Moreover, \ct{the} CNN is used for \hadi{the} CU split decision to make a comparison. Residual of \ct{the} CTU, which is obtained by pre-coding the frame using the standard HEVC, is fed into \ct{the} CNN. Pre-convolution layers in \ct{the} CNN take the residual and transform it into three depth levels, \ie $64\times64$, $32\times32$, and $16\times16$. After this operation, convolution layers extract features in all levels \ct{and} then features are concatenated and passed through fully connected layers. Finally, \ct{the} CU split decision is predicted at each depth level. \ct{The} CNN is trained offline with data generated using the HEVC encoder for several videos in different QPs and resolutions. \ct{The} CNN is preferred over SVM since it gave a better performance during the experiments.
}

\revision{Çetinkaya~\etal~\cite{VCIP20} propose a fast multi-rate encoding method using machine learning (FaMe-ML) with a specific focus on parallel encoding. The lowest quality representation is chosen as the reference. The encoding information from the reference representation and the Y, U, and V information from the raw video is fed into a convolutional neural network (CNN) to obtain a split decision for a given CTU in the given quality level. The decision from the CNN is used to speed up the parallel encoding.}

\revision{Kuang~\etal~\cite{DeepSCC} introduce a CNN-based fast prediction network for HEVC SCC. The proposed CNN structure consists of two parts where the first one is used for dynamic CTUs (\ie different content in adjacent frames), and the second one is used for stationary CTUs (\ie same content in adjacent frames). The CNN takes luminance component of the CTU and outputs the predicted mode (\ie \textit{Allskip}, \textit{Intra}, \textit{IBC}, and \textit{PLT}) for each CU inside the CTU.}

Several approaches focus on hardware implementations of the encoder using CNNs. \ekrem{Liu~\etal~\cite{ML_HardwareCNN} use a CNN to reduce CU depth search complexity in the hardware implementation.} A subsampling operation to CUs is applied, and they are given to a CNN as $8\times8$ inputs. \ct{The} CNN then predicts the best CU/PU candidate pairs but only CU texture information is used as the feature. 

Another CNN-based approach that aims to reduce complexity in the hardware implementation \ekrem{is proposed by Li~\etal~\cite{ML_HarwareTripleCNN}. As the starting point, homogeneity detection for \hadi{a} CU is done using edge strength (Sobel), QP, and motion vectors.} If the CU is determined as homogeneous, then the search is stopped, otherwise the final decision is given by \ct{the} CNN. Three sub-networks are used for \ct{the} CNN, and their outputs are combined to provide the final decision. The first network is trained with the luminance of CTU, which analyzes texture complexity. The second network is trained with Integer Motion Estimation (IME) that explores residual features of \ct{the} IME. Finally, a smaller network that is trained with motion vectors is used to analyze motion vectors. After that, \ct{the} outputs are concatenated and passed through multi-layer perceptron (MLP) layers that give a binary output that defines whether or not to split further for the given CU.

\subsubsection{Reinforcement Learning}
There have been some attempts to use Reinforcement Learning (RL) approaches in the literature as well. In RL, the main goal for the agent is to maximize reward by optimizing the actions that are taken in given situations. The environment in RL problems is modeled as Markov Decision Process (MDP). In MDP, the environment is assumed to satisfy Markov Property, meaning that the future state is only dependent on the present state and it is independent of past states.

\ekrem{Chung~\etal~\cite{ML_QRL} use a deep RL approach for CTU partitioning.} A deep Q-learning method that uses a CNN is used to predict CU split decision. The state for the algorithm is the luminance value and QP parameter, the action is split decision, and the reward is minimizing RD-cost distortion. Different models are used for each depth level, and training is done sequentially from depth $2$ towards depth $0$. 

\ekrem{Li~\etal~\cite{ML_RLMDP} model \ct{the} CU decision problem as an MDP and use an actor-critic neural network to determine early CU termination.} \ct{The} CU trajectory is used as a feature, and also neighboring CTU information is incorporated. A single model is used for different depth levels that take the action as the split decision for early termination. 

\subsection{Others}
\ekrem{Several other machine learning approaches are proposed in the literature that do not fit well into the aforementioned categories \ct{and, thus,} those methods are presented here.}

\ekrem{Huang~\etal~\cite{ML_GrayML} achieve efficient CU and PU detection by using a neural network (NN) and gray-level co-occurrence matrix (GLCM).} This approach mainly focuses on HEVC-SCC. \ekrem{A simple NN is used for each CU depth level to classify the CUs into screen content CU (SCCU) or camera-captured content CU (CCCU).} Texture complexity based on luminance values, corner point ratio, and distribution of luminance values is used as features. After classifying the CU, efficient PU mode is assigned based on statistical properties. Finally, efficient CU size is decided using information from neighboring CTUs and GLCM. GLCM is used to evaluate the texture complexity of CUs and early terminate CU search. Angular second moment (ASM) of GLCM, which represents uniformity of gray-level distribution, is used to measure the texture complexity of CU. If ASMs for both CUs are similar, they are assumed to have the same CU depth, and the search is stopped. Each neighboring CTU contributes differently to the calculation, and the weights are determined by Pearson's correlation coefficient of ASMs. 

A neural network is used for the CTU partitioning for HEVC-SCC \ekrem{by Duanmu~\etal~\cite{ML_SCCNN}.} A new feature, \ie sub-CU major directional inconsistency (MDI), which measures the consistency of features in sub-CUs, is defined. \ekrem{Separate} NN classifiers are trained offline for different QP values and CU depths using the following features: \textit{(i)} CU variance, \textit{(ii)} CU distinct color number, \textit{(iii)} CU color and gradient histogram kurtosis, \textit{(iv)} CU edge pixel percentage, and \textit{(v)} MDI of these features. \ct{The} NN outputs a number between $0-1$ rather than making a binary classification. Thus, the output value is also used as a confidence value. If it is above $0.7$, then the CU is split, if it is below $0.3$, \ct{the} CU is not split, and if it is between $0.3$ and $0.7$, full RDO is applied. 

\ekrem{Tun~\etal~\cite{ML_Genetic} model \ct{the} CTU partitioning as an optimization problem and use the Genetic Algorithm (GA) to decide on the CTU partitioning.} RD-cost based fitness function is used for \ct{the} GA and \ct{the} optimization is done based on that. To reduce the complexity of calculating \ct{the} RD-cost, \ct{the} spatial correlation between consecutive frames are used. \ct{The} keyframes are chosen periodically, and \ct{the} CTU partitioning of keyframe CTUs are shared among three consecutive frames and are used in the CTU partitioning prediction. \ct{The} optimization is stopped if the best population is not changed two times. 

Statistical analysis of image content is used as a feature and a $k$-nearest neighbors (k-NN) classifier is used in~\cite{ML_KNN}. Image complexity measure called as edge magnitude (EM) is calculated using Sobel operator \ekrem{with the formula shown in Eq.~\ref{eq:knn}.}
\begin{equation}
\label{eq:knn}
    G_{f, c} = \frac{1}{N^2}\sum_{x=0}^{N-1}\sum_{y=0}^{N-1}\sqrt{G_{h}(x, y)^2 + G_{v}(x,y)^2}
\end{equation}
where $N$ is \ct{the} width of \ct{the} CU, $f$ is \ct{the} frame number, $c$ is \ct{the} CU address, $G_{h}$ and $G_{v}$ are \ct{the} horizontal and vertical gradients computed by \ct{the} Sobel operator at pixel location $(x,y)$. \ct{The} EM is then used for an early split decision. Moreover, CU splitting is modeled as a two-class classification problem and Fischer's linear discriminant analysis (FLDA) is used to transform statistical data into a more separable format. FLDA is used for reducing \ct{the} dimensionality in the data \ct{in order to} become easier for \ct{the} classification. The k-NN classifier is then used to directly estimate the depth of \ct{the} given CU. If \ct{the} RD cost and \ct{the} EM are high, \ct{the} CU is classified into split class. Moreover, statistical parameters are updated online when there is a scene change or fast movement using training frames which are determined by the edge magnitude ratio (ER) between frames.

All machine learning based approaches and commonly used features are summarized in Table~\ref{table:ML}.

\begin{table*}[pos=t]
\centering
\caption{Machine Learning Based Approaches}
\label{table:ML}
\resizebox{\textwidth}{!}{%
\begin{tabular}{l|l|l|l}
\multicolumn{1}{c|}{\textbf{Method}} & \multicolumn{1}{c|}{\textbf{Approach}} & \multicolumn{1}{c|}{\textbf{Features}}                                             & \multicolumn{1}{c|}{\textbf{Mode}} \\ \hline
\cite{ML_SVM}                        & SVM                                    & MSE, Number of Encoded Bits (NEB)                                                  & Inter                              \\
\cite{ML_WSVM}                       & SVM                                    & Depth Values, RD Cost, Texture Complexity                                          & Inter                              \\
\cite{ML_FlexSVM}                    & SVM                                    & Depth Values, skip Mode, Motion Vectors, QP                                        & Intra                              \\
\cite{ML_FlexSVMImp}                 & SVM                                    & Variance, Texture Complexity                                                       & Intra                              \\
\cite{ML_LSVM}                       & SVM, Neighbour CTU                     & Depth Values, Texture Complexity, Hadamard Cost                                    & Intra                              \\
\cite{ML_ANNSVM}                     & SVM                                    & Depth Values, Variance                                                             & Intra                              \\
\cite{ML_FuzzySVM}                   & SVM, Neighbour CTU                     & Depth Values, SAD, RD Cost, Motion Vectors, QP                                     & Inter                              \\
\cite{ML_DoubleSVM}                  & SVM, Neighbour CTU                     & Depth Values, Motion Vectors, Texture Complexity, RD Cost                          & Intra                              \\
\cite{ML_TwoLayerSVM}                & SVM                                    & Depth Values, Texture Complexity, RD Cost                                          & Intra                              \\
\cite{ML_TwoLayerSVMInter}           & SVM, Neighbour CTU, Co-located CTU     & Depth Values, Texture Complexity, RD Cost, QP                                      & Inter                              \\
\cite{ML_SVMSCC}                     & SVM, Neighbour CTU                     & Depth Values, Luminance Values of CTU, Bitrate, RD Cost, QP, SCC Mode             & Intra                              \\
\cite{ML_DecisionTree}               & Decision Tree                          & Sub-CU Difference, Variance, Edge Complexity, Pixel Values                         & Intra                              \\
\cite{ML_DataTree}                   & Decision Tree                          & DCT Coefficients, Variance                                                         & Intra                              \\
\cite{ML_DecisionSCC}                & Decision Tree, HEVC SCC           & Variance, Luminance Values of CTU, Edge Complexity, SCC Mode, RD Cost              & Intra                              \\
\cite{DecTree3D} & Decision Tree, 3D-HEVC & Variance, SAD, Texture & Intra \\
\cite{Overview3D} & Decision Tree, 3D-HEVC & DIS Cost, Depth Values, QP & Intra \\
\cite{ML_RForest}                    & RF, Neighbour CTU                      & Luminance Values of CTU                                                            & Intra                              \\
\cite{ML_RForestFeatureExtr}         & RF                                     & Depth Values, Variance, Pixel Values, Motion Vectors, RD Cost, Edge Complexity, QP & Inter/Intra                        \\
\cite{MR_MLBased}                    & RF, Multirate                         & Depth Values, Variance, Motion Vectors                                             & Intra                              \\
\cite{MR_RFHBR}                      & RF, Multirate                         & Depth Values, QP                                                                   & Inter                              \\
\cite{ML_BayesianOnline}             & Bayesian                               & Depth Values, Scene Change                                                         & Inter/Intra                        \\
\cite{ML_BayesianOnlineGMM}          & Bayesian                               & Depth Values, Scene Change,  Luminance Values                                      & Inter/Intra                        \\
\cite{ML_BayesianProgressive}        & Bayesian, Neighbour CTU                & Depth Values, Variance, Texture Complexity                                         & Intra                              \\
\cite{ML_BayesianSCC}                & Bayesian                               & RD Cost, QP, SCC Mode                                                              & Intra                              \\
\cite{ML_FastCuNN}                   & Neural Network                         & Depth Values, Motion Vectors                                                       & Inter/Intra                        \\
\cite{ML_LSTMCNN}                    & CNN, LSTM, Co-located CTU              & Depth Values, Pixel Values                                                         & Inter/Intra                        \\
\cite{ML_DeepCNN, ML_AsymCNN}        & CNN                                    & Luminance Values of CTU                                                            & Intra                              \\
\cite{ML_TextureCNNMix}              & CNN                                    & Depth Values, QP, Texture Complexity, Luminance Values of CTU                      & Intra                              \\
\cite{ML_TwoCNNCUPU, ML_HardwareCNN} & CNN                                    & Luminance Values of CTU, QP                                                        & Intra                              \\
\cite{ML_CNNROI_2}                   & CNN                                    & Luminance Values of Frame                                                          & Intra                              \\
\cite{ML_SVMCNN}                     & CNN, SVM                               & Luminance Values of Frame                                                          & Inter                              \\
\cite{VCIP20}                                & CNN, Multirate                                        & Depth Values, Reference Frames, Prediction Mode, RD Cost, Variance                               & Inter                                   
\\
\cite{DeepSCC}                                & CNN, HEVC SCC                                       & Luminance Values                              & Intra         
\\
\cite{ML_HarwareTripleCNN}           & CNN                                    & Luminance Values of CTU, Edge Complexity, QP, Motion Vectors                       & Inter                              \\
\cite{ML_QRL}                        & RL, CNN                                & Luminance Values of CTU, QP                                                        & Intra                              \\
\cite{ML_RLMDP}                      & RL, Neural Network, Neighbour CTU      & Depth Values                                                                       & Inter                              \\
\cite{ML_GrayML}                     & Neural Network, Neighbour CTU          & Texture Complexity, Luminance Values of CTU                                        & Intra                              \\
\cite{ML_SCCNN}                      & Neural Network                         & Variance, CTU Color Information, Edge Complexity                                   & Intra                              \\
\cite{ML_Genetic}                    & Genetic Algorithm, Frame History       & Depth Values                                                                       & Inter                              \\
\cite{ML_KNN}                        & KNN Classifier                         & Texture Complexity, Edge Complexity                                                & Intra                             
\end{tabular}
}
\end{table*}

\subsection{Summary}

Different machine learning approaches have been proposed to \ekrem{tackle} the CTU depth decision problem. The main trade-off for ML based approaches is the trade-off between increased accuracy and time complexity. SVMs and RFs are still the main traditional approaches due to their low\hadi{er} complexity and simplicity. There still exists recent works that benefit from them \cite{ML_FuzzySVM, ML_DoubleSVM, ML_TwoLayerSVM, ML_SVMSCC, ML_DecisionSCC, ML_RForestFeatureExtr, MR_RFHBR}. \ct{Recently, multiple SVMs and RFs have been proposed more commonly}~\cite{ML_FlexSVMImp, ML_LSVM, ML_ANNSVM, ML_FuzzySVM, ML_DoubleSVM, ML_TwoLayerSVM, ML_SVMSCC, ML_DecisionSCC, ML_RForestFeatureExtr}. This allows each SVM or RF to focus on a specific feature which results in a better prediction and generalization. The most common features used for SVM and RF-based methods are \textit{(i)} CU depth values, \textit{(ii)} texture complexity, and \textit{(iii)} RD cost. In Bayesian learning methods, features to update statistical properties are also chosen manually, and depth values are utilized in all of the Bayesian learning approaches since they are important. \ct{Additionally}, scene change detection is key in deciding when to update the statistics so that different approaches are chosen to decide on the correct update period, \ie online learning. Features that are used by ML methods are given in Table \ref{table:ML}.

One downside of the traditional ML methods is that the features need to be selected manually. This requires expert-level domain knowledge since the performance of the technique heavily depends on the feature set. This is not wanted in many cases since manually crafting features is prone to errors, and it can also introduce human bias to the ML method. The main advantage of traditional ML methods is \ct{that} they do not introduce significant time complexity. However, their performance is not on par with their deep learning based counterparts. Giving all the available data to the ML method and allowing it to extract useful features from the data usually results in better performance, and deep learning methods are an excellent example of that. In almost all applications, deep learning methods give better performance compared to \hadi{the} traditional ML methods~\cite{dncnn, objectdetectsurvey, actionrecog}. Also, deep learning methods are better at utilizing the excess amount of available data, which is nowadays accessible for almost all applications.

Deep learning methods, on the other hand, impose a significant time complexity to the encoder, which is usually not desired. CNNs are used commonly due to the nature of the videos, which consist of consecutive 2D data. We can see the majority of approaches use the luminance value of \ct{the} CTU as the main data for feeding the CNN and let the network extract useful features from these values. Some approaches also append additional features along with luminance values to feed the network~\cite{ML_FastCuNN, ML_TwoCNNCUPU}. \ct{Furthermore}, we can see that some methods try to solve \ct{the} CTU partitioning in a top-down approach and run a CNN for each depth level~\cite{ML_FastCuNN, ML_DeepCNN, ML_AsymCNN, ML_TextureCNNMix} while others try to predict the whole CTU partitioning in one step to minimize the time-complexity~\cite{ML_LSTMCNN}. Since CTU partitioning can be modeled as MDP, several RL-based approaches are also proposed~\cite{ML_QRL, ML_RLMDP}. We can still observe CNNs are the critical factors in RL-based methods~\cite{ML_QRL}. 

In addition to \textit{(i)} traditional, \textit{(ii)} deep learning, and \textit{(iii)} RL based methods, several different ML approaches, \ekrem{\ie kNN classifier, neural network, and genetic algorithm,} are also used~\cite{ML_GrayML, ML_SCCNN, ML_Genetic, ML_KNN}. We can categorize these methods in the traditional ML approaches, but we believe they are in between traditional and deep learning based methods and, thus, we decided to categorize them separately. There have also been some attempts that tried to combine both statistic based and ML based approaches to benefit from them at the same time~\cite{ML_ANNSVM, ML_TextureCNNMix, ML_RLMDP}.

\begin{figure*}[pos=t]
    \captionsetup{justification=centering,
                  singlelinecheck=false,
                  font=sf, labelfont=bf} 
    \centering
    \begin{subfigure}{0.49\textwidth}
    \includegraphics[width=\textwidth,trim={1cm 0.5cm 1cm 1cm},clip]{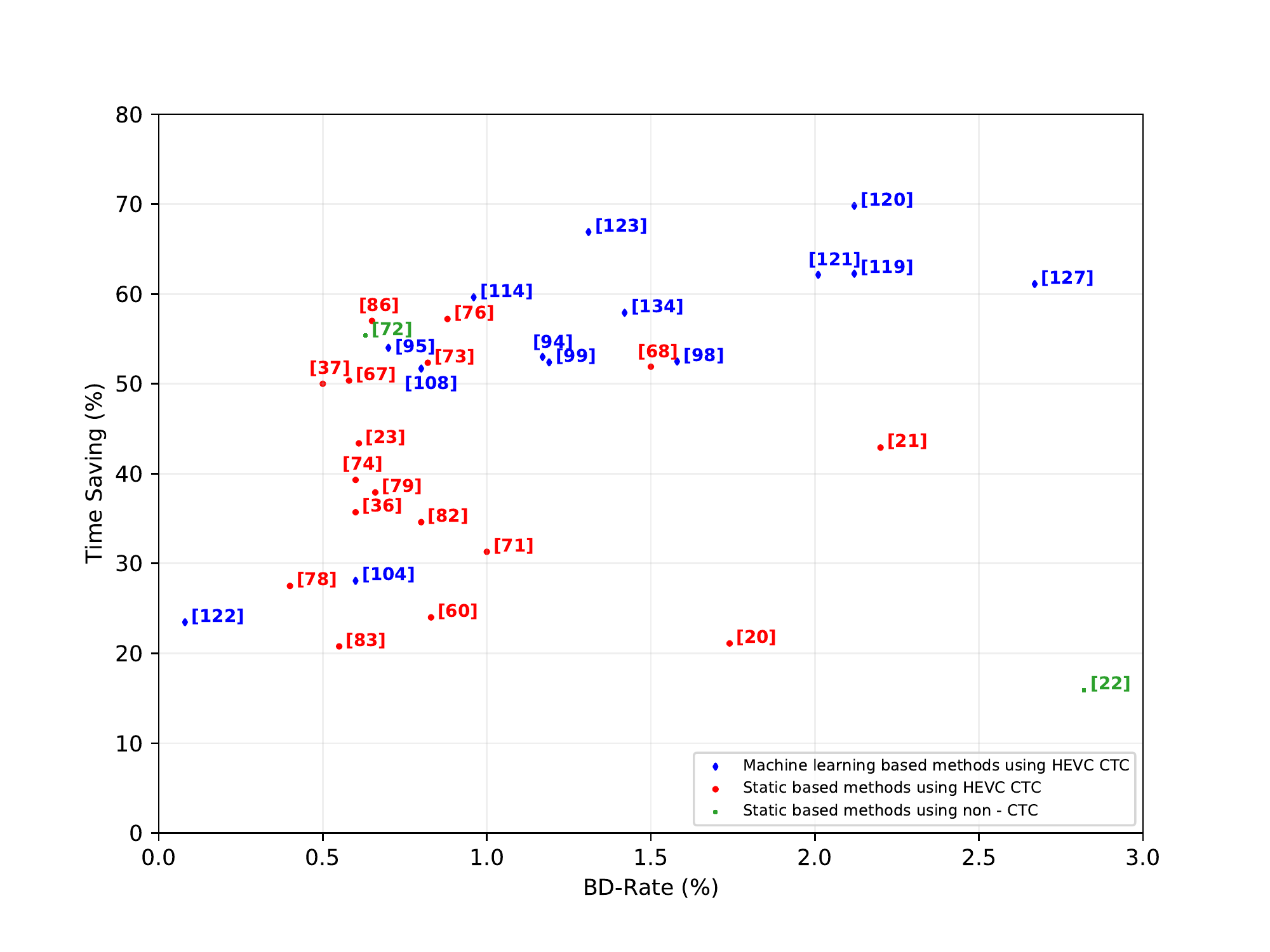}
    \caption{Intra coding approaches}
    \end{subfigure}
    \hfill 
    \begin{subfigure}{0.49\textwidth}
    \includegraphics[width=\textwidth,trim={1cm 0.5cm 1cm 1cm},clip]{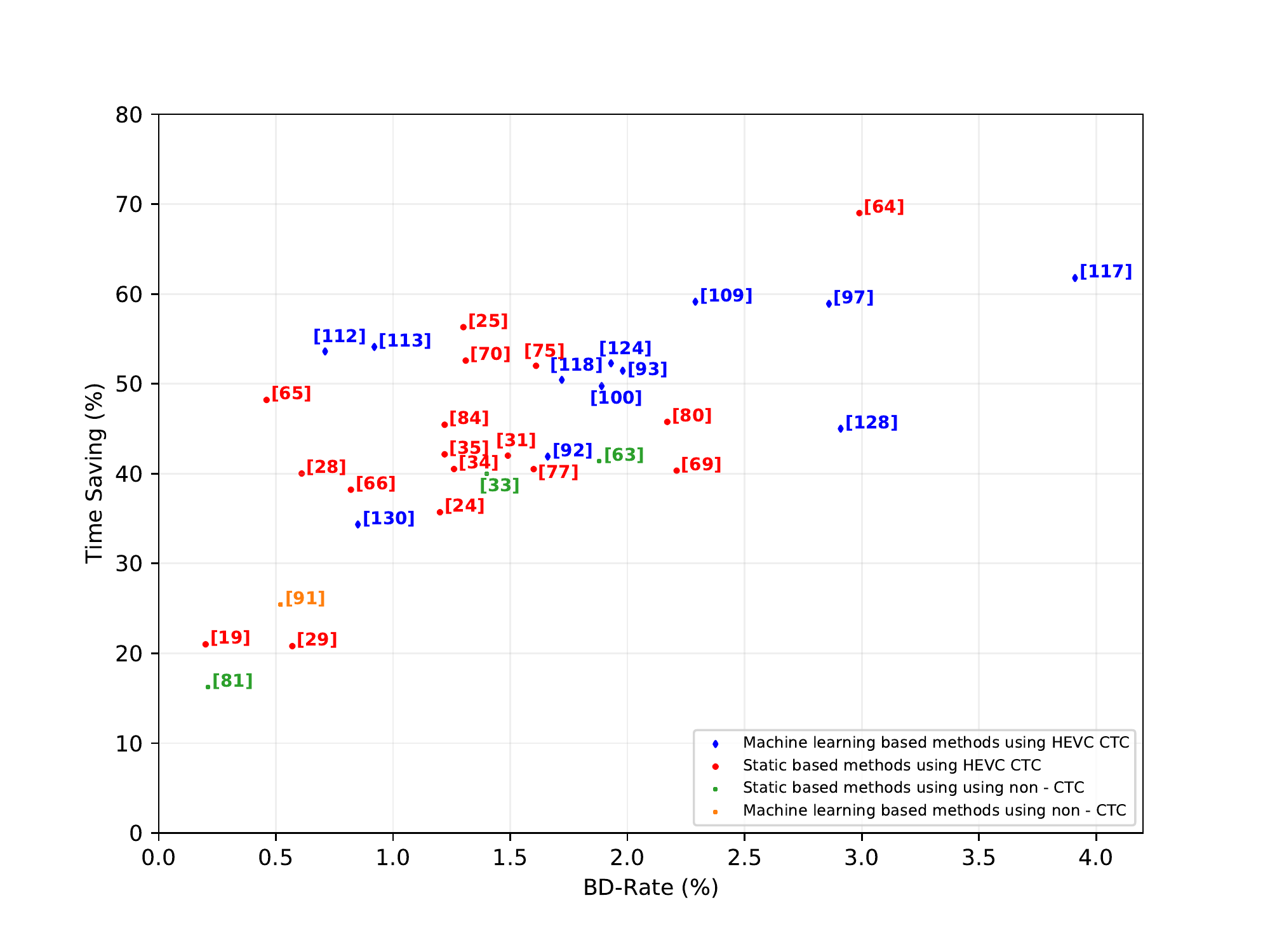}
    \caption{Inter coding approaches}
    \end{subfigure}
    \caption{\revision{Encoding time-saving vs. BD-Rate graphs of CTU partitioning methods for the base HEVC. Methods are categorized into statistics (Stat) and ML-based (ML) approaches along with the evaluation set type. }}
    \label{fig:rdcosts}
\end{figure*}

\revision{Encoding time-saving vs. BD-Rate graphs for both inter and intra coding approaches are depicted in Fig.~\ref{fig:rdcosts}. Methods that use HEVC extensions are omitted. Approaches are categorized into statistics-based and ML-based methods. Moreover, the used test sequences is used as another label to make a fair comparison between approaches. Approaches that have used sequences from HEVC Common Test Conditions~\cite{HEVC_CTC} are labeled with CTC, otherwise, with non-CTC.
All results are obtained from the reported results in the corresponding papers.}


\section{Discussion and Future Work}
\label{sec:discussion}
There has been an increase in machine learning based approaches in \ct{the} last years. However, statistics based approaches are still applied in the literature. The main problem with ML based approaches is the complexity introduced by the ML method resulting in simple network structures being used so far. However, recent developments in deep learning suggest that using deeper convolutional networks significantly improves \ct{the} performance, specifically while working with image data since deeper networks are better at capturing hierarchical information, which is present in many situations~\cite{Inception, Residual, PowerofDepth}. Moreover, recent advances in the deep learning field made it possible to use very complex models with significantly reduced inference time even for challenging tasks \cite{YOLOv2, YOLOv3}. Moreover, to overcome substantial data requirements of deep learning methods, few-shot or one-shot learning methods have been proposed, and their performance had improved substantially~\cite{FewShotModelAgnostic, FewShotLearning, OneShotMatching}. Furthermore, these methods are combined and used in video-related tasks, which is challenging for deep learning techniques, and resulted in a good performance despite using little data \cite{OneShotMixVideo}. We believe a few-shot learning approach can be helpful in the CTU partitioning decisions since each video has different characteristics. The better approach will be tailoring the model for each video, which might be possible with this improved training scheme. Traditional ML methods are still helpful in cases where the task is evident. Thus, it is easy to design features, and the available data is not sufficient enough to fully utilize a deep learning method. However, we believe the main direction to follow is to preserve the performance of deep learning methods and decrease their complexity. 

On the other hand, for statistics based approaches, unexplored statistical correlations can still be found, and existing correlations can be benefited better. The main advantage of statistical approaches is that they do not introduce any noticeable time-complexity in the encoding process. One possible approach might be to use ML methods to extract statistical information beforehand to have a better understanding and move the complexity of ML out of the encoding framework. 

{Emerging video codecs like AV1~\cite{av1} and Versatile Video Coding (VVC)~\cite{vvc} introduce more bitrate reduction compared to HEVC~\cite{HEVC}, while their encoding time complexity increases which makes their usage costly and challenging, especially for online applications.}

\begin{figure}[pos=t]
    \captionsetup{justification=centering,
                  singlelinecheck=false,
                  font=sf, labelfont=bf} 
    \centering
    \includegraphics[width=\linewidth]{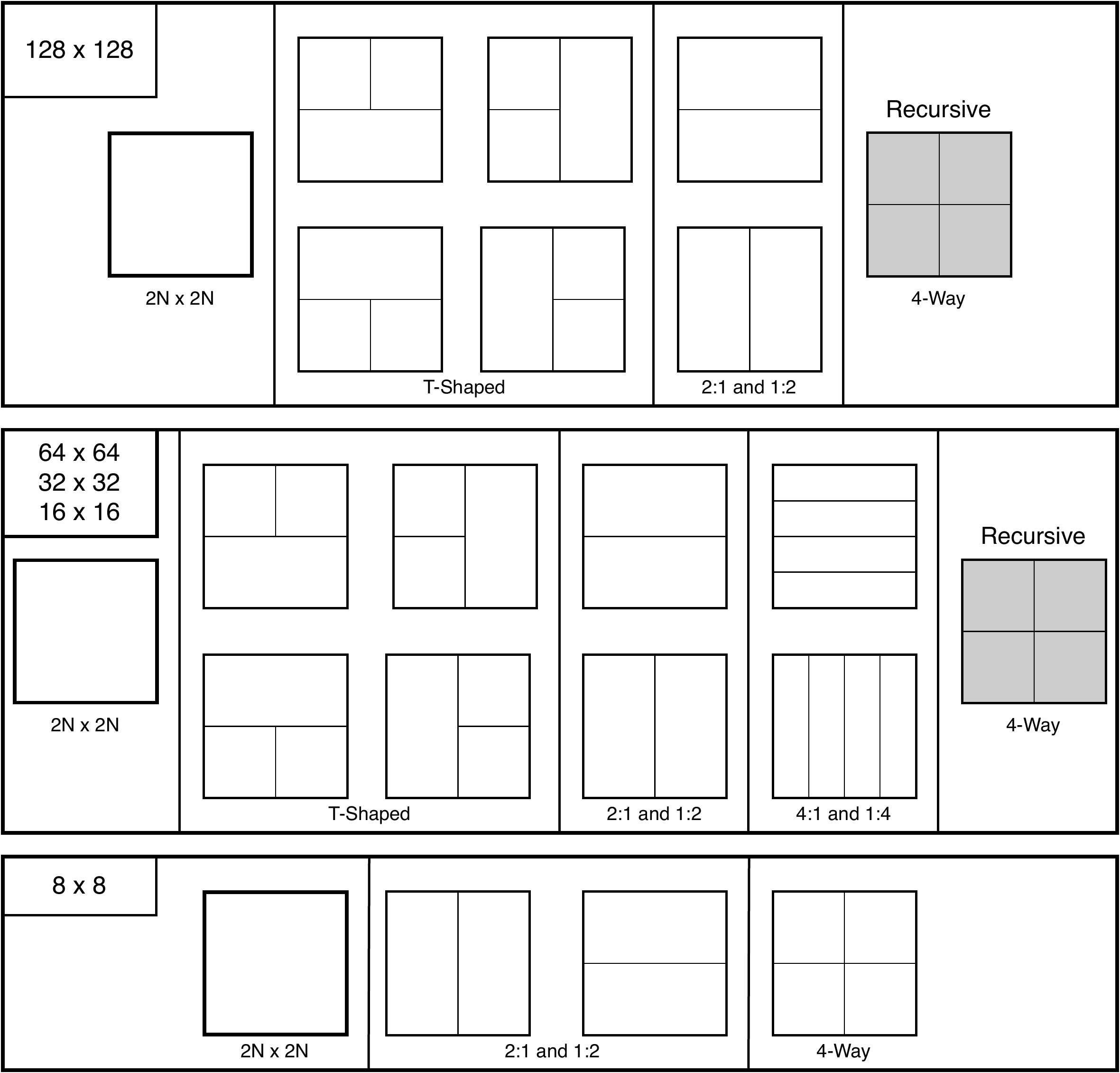}
    \caption{AV1 superblock partitioning.}
    \label{fig:av1_partition}
\end{figure}

AV1 and VVC introduce different block partitioning structures than HEVC.
\ekrem{AV1 uses superblock partitioning that starts from $128\times 128$ sized blocks. A 10-way partition-tree structure is used to sub-partition each block. For each $2N\times2N$ block there exist \begin{enumerate*}[label=\textit{\alph*)}]
\item one $2N\times2N$,
\item four "T" shaped,
\item two 2:1 and 1:2,
\item two 4:1 and 1:4, and
\item one 4-way
\end{enumerate*} split patterns. \ekrem{Among the aforementioned split patterns, 4-way split is the only pattern that can be recursively partitioned until it reaches the lowest $4\times4$ size.} Moreover, there are some special cases for AV1 superblock partitioning: \begin{enumerate*}[label=\textit{\alph*)}]
\item two 4:1 and 1:4 splits are not available for $128\times128$ and $8\times8$ blocks, and
\item "T" shaped split is not available for 8×8 blocks
\end{enumerate*}. \ct{The} \ekrem{AV1 superblock partitioning} is summarized in Fig.\ref{fig:av1_partition}.}

{VVC utilizes Quaternary Tree plus (Multi) binary-ternary Tree (QTMT) coding block structure with the maximum CTU size of $128\times128$ pixels. Both binary and ternary splits are enabled for each leaf node of the quad-tree which makes the partitioning more flexible and complex compared to HEVC. Leaf nodes of multi-type trees are called Coding Units (CUs), which are not units only for coding but for prediction and transform as well.}

\begin{figure}[pos=t]
    \captionsetup{justification=centering,
                  singlelinecheck=false,
                  font=sf, labelfont=bf} 
    \centering
    \includegraphics[width=0.5\textwidth]{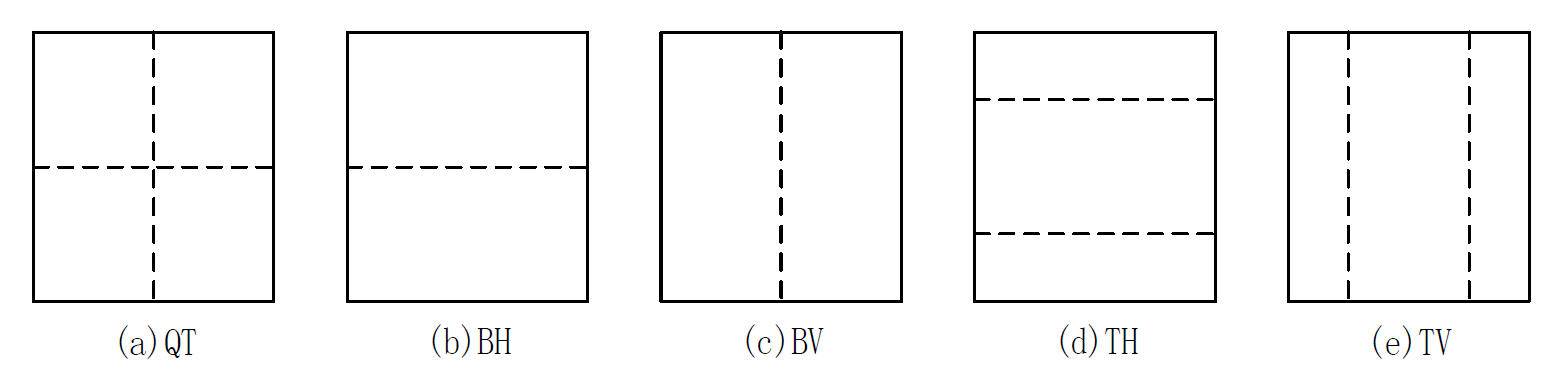}
    \caption{VVC partitioning.}
    \label{fig:QTMT}
\end{figure}

\ekrem{In VVC, each block can be split into five patterns: \begin{enumerate*}[label=\textit{\alph*)}]
\item one Quad-Tree (QT) structure,
\item two Binary Tree (BT) structures including horizontal binary tree (BH) and vertical binary tree (BV), and
\item two Ternary Tree (TT) structures including horizontal ternary tree (TH) and
vertical ternary tree (TV)
\end{enumerate*}. These five structures are shown in Fig~.\ref{fig:QTMT}. It should be noted that the QT structure is not allowed for non-QT splits.}

\revision{On top of these structures, VVC includes some additional partitioning schemes as well. \textit{Geometric partitioning mode} (GPM)~\cite{GPM_VVC} is introduced to increase the partitioning precision of moving natural objects. This is achieved by utilizing non-rectangular and asymmetric partitioning on top of the existing block partitioning structure of the VVC. By allowing non-rectangular and asymmetric splitting of the blocks, VVC with GPM can better approximate the curved edges which are difficult to estimate using only rectangular blocks.}

\revision{In addition to the wide angle intra prediction and other sophisticated tools in VVC compared to HEVC, the Intra Subpartition Coding (ISP) coding mode was introduced to further improve the intra prediction coding efficiency. In the ISP coding mode, luma intra-predicted blocks are enabled to be divided into horizontally or vertically equal size subpartitions that have at least 16 samples. For example,  a $16\times32$ pixels input block can be divided into four $4\times32$ pixels or four $16\times8$ pixels subblocks.}

\revision{Affine motion estimation is another sophisticated tool that has been added to the VVC to extend the classical translation (2 DOF) to more degrees of freedom to include rotation, scaling, aspect ratio, and shearing translations~\cite{affine}.}

\ekrem{Although VVC and AV1 achieve efficient complexity reduction, there is only a tiny number of improved CTU partitioning methods available in the literature due to their recent developments. However, since both of these codecs follow a block structure architecture, algorithms introduced in this survey can be implemented in these codecs either directly or with minor modifications.}

\ekrem{Gu~and~Wen~\cite{av1-2} propose a mid-depth based block structure determination for AV1 which is similar to~\cite{H_FastPyramid} and~\cite{av1-hevc} combined.}

\ekrem{Chen~\etal~\cite{vvc2} propose an algorithm for VVC that consists of three steps. First, \ct{the} variance of pixels for $32\times32$ CUs is calculated, and if it is lower than a pre-determined threshold, \ct{the} CU is classified to be homogeneous, and further splitting is stopped. Second, absolute gradients of each pixel in horizontal and vertical directions are computed using the Sobel operator ($D_x, D_y$). If \ekrem{conditions in Eq.~\ref{eq:vvc2}} are met for $32\times32$ CU, it is partitioned by QT structure and other options are skipped:}

\begin{equation}
\label{eq:vvc2}
\begin{split}
    ((1<D_x/D_y<TH_2)~or~(1<D_y/D_x<TH_2))
    \\and~((D_x>TH_3)~and~(D_y>TH_3))
\end{split}
\end{equation}

\ekrem{Finally, if neither of the aforementioned conditions are met, the variance of each sub-block is computed for all five types of splitting and then partitioning is continued only for the type that has the maximum variance.}

\revision{Bordes~\etal~\cite{JVET_J0022} use a set of fast methods to speed up the encoding process in JEM, the predecessor of VVC that was an experimental software codebase based on the HEVC. Caching mechanism are introduced to reduce redundant computations. Various RD-cost based heuristics were also introduced to reduce the testing combinatory.}

\revision{Galpin~\etal~\cite{JVET_J0034} proposes a CNN-based approach for speeding up VVC Intra slicing. The texture complexity of the block is analyzed using a CNN, which predicts the most probable splits inside each potential sub-block. Two CNNs are used separately, one for the luma component and the other for the chroma components. Output probabilities from the CNN are averaged for each split boundary in the block, and the decision to whether to search for or skip each split option is given based on the pre-determined threshold values.}

\section{Conclusion}
\label{sec:conclusion}
In this paper, existing CTU partitioning approaches for HEVC are surveyed. \ct{The m}ethods are categorized into two major groups: \textit{statistics} based and \textit{machine learning} based. In statistic based approaches, \ct{the} proposed methods exploit statistical similarity in the video by manually defining a set of features, thresholds, rules, \etc~Neighboring approaches use the available information in the spatially and temporally neighboring CTUs. In contrast, inherent approaches focus on information available within the current CTU. Similar to the statistics based approaches, machine learning based approaches exploit similarity in the video but benefit from machine learning methods during the process. Traditional ML methods such as SVM and RF are used for which the features need to be chosen manually. Deep learning based methods, on the other hand, extract useful features implicitly. \ekrem{Recent emerging codecs, AV1 and VVC, provide improved encoding structure using new tools and improving existing ones. However, due to their recent development, there are not enough studies conducted for block partitioning of these codecs. }

\section*{Acknowledgment}

The financial support of the Austrian Federal Ministry for Digital and Economic Affairs, the National Foundation for Research, Technology and Development, and the Christian Doppler Research Association, is gratefully acknowledged. Christian Doppler Laboratory ATHENA: \url{https://athena.itec.aau.at/}.

\balance
\bibliographystyle{model1-num-names.bst}
\bibliography{bibtex/bib/main}

\end{document}